\documentclass{article}
\usepackage{graphicx}
\usepackage{amsmath}
\usepackage{amsfonts}
\usepackage{amssymb}

\begin{document}

\thispagestyle{empty}
\hfill{{\protect\footnotesize CBPF-NF-018/03}}

\begin{center}

{\Large\bf Pascual Jordan, his contributions to quantum mechanics and his legacy in
contemporary local quantum physics}

\vspace{0.3cm}

{\sl Bert Schroer 

\vspace{0.3cm}

present address: CBPF, Rua Dr. Xavier Sigaud 150,\\[-1mm] 
22290-180 Rio de Janeiro, Brazil\\[-1mm]
email schroer@cbpf.br\\[-1mm]
permanent address: Institut für Theoretische Physik\\[-1mm]
FU-Berlin, Arnimallee 14, 14195 Berlin, Germany}

May 2003

\end{center}

\begin{abstract}
After recalling episodes from Pascual Jordan's biography including his pivotal
role in the shaping of quantum field theory and his much criticized conduct
during the NS regime, I draw attention to his presentation of the first phase
of development of quantum field theory in a talk presented at the 1929 Kharkov
conference. He starts by giving a comprehensive account of the beginnings of
quantum theory, emphasising that particle-like properties arise as a
consequence of treating wave-motions quantum-mechanically. He then goes on to
his recent discovery of quantization of ``wave fields'' and problems of gauge
invariance. The most surprising aspect of Jordan's presentation is however his
strong belief that his field quantization is a transitory not yet optimal
formulation of the principles underlying causal, local quantum physics. The
expectation of a future more radical change coming from the main architect of
field quantization already shortly after his discovery is certainly quite startling.

I try to answer the question to what extent Jordan's 1929 expectations have
been vindicated. The larger part of the present essay consists in arguing that
Jordan's plea for a formulation without ``classical correspondence crutches'',
i.e. for an intrinsic approach (which avoids classical fields altogether), is
successfully addressed in past and recent publications on local quantum physics.
\end{abstract}

\newpage
\setcounter{page}{1}

\section{Biographical Notes}

There are not many physicists in whose biography the contradictions of human
existence, the proximity of glorious scientific achievements and disturbing
human weaknesses in the face of the great catastrophe of the 20th century, are
as starkly reflected as in the personality of Pascual Jordan\footnote{The
original title ``Pascual Jordan, Glory and...'' has been changed since
although the birth of quantum theory represents one of the most glorious
epochs in physics, Jordan's himself remained ``the unsung hero'' among the
creators of that theory \cite{Schweber}.}.

Born on October 18, 1902 in Hannover of mixed German-Spanish ancestry, he
became (starting at age 22) a main architect of the conceptual and
mathematical foundations of quantum theory and the protagonist of quantum
field theory. Pascual Jordan owes his Spanish name to his great grandfather
Pascual Jorda (probably unrelated to the biblical river), who came from the
Alcoy branch (southern Spain) of the noble Jorda family with a genealogy which
can be traced back to the 9th century. After the British-Spanish victory of
Wellington over Napoleon, the family patriarch Pascual Jorda settled in
Hannover where he continued his service to the British crown as a member of
the ``Koeniglich-Grossbritannisch-Hannoverschen Garde-Husaren Regiments''
until 1833. Every first-born son of the Jordan (the n was added later) clan
was called Pascual.

There is no doubt that Pascual Jordan took the lead in the formulation of the
conceptual and mathematical underpinnings of ``Matrix Mechanics'' in his
important paper together with Max Born \cite{B-J} submitted on 27. September
1925 (3 months after the submission of Heisenberg's pivotal paper!) entitled
``Zur Quantenmechanik''. His mathematical preparation, particularly in the
area of algebra, was superb. He had taken courses at the G\"{o}ttingen
mathematics department from Richard Courant and became his assistant (helping
in particular on the famous Courant-Hilbert book); through Courant he got to
know Hilbert before he met the 20 year older Max Born, the director of the
Theoretical Physics department of the G\"{o}ttingen university. By that time
Jordan already had gained his physics credentials as a co-author of a book
which he was writing together with James Franck \cite{Franck}.

After Max Born obtained Heisenberg's manuscript, he tried to make sense of the
new quantum objects introduced therein. While he had the right intuition about
their relation to matrices, he felt that it would be a good idea to look for a
younger collaborator with a strong mathematics background. After Pauli
rejected his proposal (he even expressed some reservations that Born's more
mathematically inclined program could stifle Heisenberg's powerful physical
intuition), Jordan volunteered to collaborate in this problem \cite{Pais}%
\cite{Jammer}. Within a matter of days he confirmed that Born's conjecture was
indeed consistent. The Born-Jordan results made Heisenberg's ideas more
concrete. Probably as a consequence of the acoustic similarity of pq with
Pascual, the younger members of the physics department (the protagonists of
the ``Knabenphysik'') in their discussions often called it the Jordan
relation. Max Born became Jordan's mentor in physics. Jordan always maintained
the greatest respect which withstood all later political and ideological differences.

The year 1925 was a bright start for the 22-year-old Jordan. After the
submission of the joint work with Max Born on matrix mechanics, in which the
p-q commutation relation appeared for the first time, there came the famous
''Dreimaennerarbeit''\cite{B-H-J} with Born and Heisenberg in November of the
same year, only to conclude the year's harvest with a paper by him alone on
the ``Pauli statistics''. Jordan's manuscript contained what is nowadays known
as the Fermi-Dirac statistics; however it encountered an extremely unfortunate
fate after its submission because it landed on the bottom of one of Max Born's
suitcases (in his role as the editor of the Zeitschrift fuer Physik) on the
eve of an extended lecture tour to the US, where it remained for about half a
year. When Born discovered this mishap, the papers of Dirac and Fermi were
already in the process of being published. In the words of Max Born
\cite{Born}\cite{Schucking} a quarter of a century later: ''I hate Jordan's
politics, but I can never undo what I did to him......When I returned to
Germany half a year later I found the paper on the bottom of my suitcase. It
contained what one calls nowadays the Fermi-Dirac statistics. In the meantime
it was independently discovered by Enrico Fermi and Paul Dirac. But Jordan was
the first''\footnote{In a correspondence with Stanley Deser, Stanley added a
light Near East touch by remarking that without Max Born's faux pas the
Fermions would have been called ``Jordanions''.}. In Jordan's subsequent
papers, including those with other authors such as Eugene Wigner and Oscar
Klein, it was always referred to as ``Pauli statistics'' because for Jordan it
resulted from a sraightforward algebraization of Pauli's exlusion principle.

>From later writings of Born and Heisenberg we also know that Jordan
contributed the sections on the statistical mechanics (or rather kinetic gas
theory) consequences to the joint papers on matrix mechanics; this is not
surprising since the main point in his 1924 PhD thesis was the treatment of
photons according to Planck's distribution whereas thermal aspects of matter
were described according to Boltzmann. He continued this line of research by
introducing the ``Stosszahlansatz'' for photons and using for electrons and
atoms the Bose statistics \cite{thermal}\footnote{This paper was submitted
simultaneously with another paper in which Jordan coined the term
``Pauli-Principle'' \cite{Pauli}; but the relation to statistics was only seen
later.} which brought praise by Einsteins and led to an unfortunately largely
lost correspondence. In the following we will continue to mention his
scientific contributions in the biographical context and reserve a more
detailed account about their scientific content to the next section.

The years 1926/27 were perhaps the most important years in Jordan's career in
which he succeeded to impress his peers with works of astonishing originality.
The key words are Transformation Theory \cite{Trans}\cite{Trans2} and
Canonical Anti-Commutation Relations \cite{Gas}. With these discoveries he
established himself as the friendly competitor of Dirac on the continental
side of the channel and in its printed form one finds an acknowledgment of
Dirac`s manuscript\footnote{In those days papers were presented in a factual
and very courteous style; however verbal discussions and correspondences were
sometimes more direct and less amiable (e.g. see some published letters of
Pauli \cite{Pais}\cite{Schweber}).}. As an interesting sideline, one notes
that in a footnote at the beginning of the paper about transformation theory
Jordan mentions a ``very clear and transparent treatment'' of the same problem
in a manuscript by Fritz London, a paper which which he received after
completing his own work and which was published in \cite{London}. So it seems
that the transformation theory was discovered almost simultaneously by three
authors. Most physicists are more familiar with Dirac's notation (as the
result of his very influential textbook whose first edition appeared in 1930).
Jordan's most seminal contribution is perhaps his 1927 discovery of
``Quantization of Wave Fields'' which marks the birth of QFT. The reader finds
a description about the chronology of this most important of of Jordan's
discoveries, its relation to Dirac's radiation theory and its influence on the
subsequent development of particle physics in the next section.

We are used to the fact that in publications in modern times the relation of
names to new concepts and formulas should be taken with a grain of salt e.g.
if we look at the original publication of Virasoro we are not terribly
surprised that the algebra has no central term and consists only of
frequencies of one sign (i.e. is only half the Witt algebra). But we trust
that what the textbooks say about the beginnings of quantum mechanics can be
taken literally. When we look at \cite{Trans} page 811 we note that the
relation between Born and the probability interpretation is much more indirect
than we hitherto believed. Born in his 1926 papers was calculating the Born
approximation of scattering theory and his proposal to associate a probability
with scattering in modern terminology concerns the interpretation of the cross
section. The generalization to the probability interpretation of the absolute
square of the x-space Schr\"{o}dinger wave function according to Jordan was
done by Pauli\footnote{This may partially explain why Pauli in his 1933
Handbuch article on wave mechanics introduced the spatial localization
probability density without reference to Born.} who was of course strongly
influenced by Born. It happened frequently that new ideas were used freely in
scientific discussions and that the later attachment of a person's name may
represent the correct origin of that mode of thinking but cannot be taken
prime facie when it comes to historical details.

His increasing detachment after 1933 from the ongoing conceptual development
of QFT, and his concentration on more mathematical and conceptual problems of
quantum theory whose investigation proceedes at a slower pace (and can be done
without being instantly connected to the stream of new information) happens at
the time of his unfortunate political activities, as he lets himself be sucked
more and more into the mud of the rising Nazi-regime\footnote{National
Socialist=NS=Nazi. The colloquial expressions as Nazi and Sozi (socialist)
originated from the times before Hitler's rise to power at the time of the big
street fights between the rightwing and leftwing rowdies. The origin of the
terminology ``Nazi'' is a bit like ``Commie'' (for communists) in the US, and
for this reason not used by historians.}. In trying to understand some of his
increasing nationalistic and militaristic behavior it appears of some help to
look at his background and upbringing, although a complete understanding would
probably escape those of us who did not experience those times of great
post-war turmoil.

Pascual Jordan was brought up in a traditional religious surrounding. At the
age of 12 he apparently went through a soul-searching fundamentalist period
(not uncommon for a bright youngster who tries to come to terms with rigid
traditions) in which he wanted to uphold a literal interpretation of the bible
against the materialistic Darwinism (which he experienced as a ''qu\"{a}lendes
Aergernis'', a painful calamity), but his more progressive teacher of religion
convinced him that there is basically no contradiction between religion and
the sciences. This then became a theme which accompanied him throughout his
life; he wrote many articles and presented innumerable talks on the subject of
religion and science.

At the times of the great discoveries in QT many of his colleagues thought
that the treaty of Versailles was unjust and endangered the young Weimar
Republic, but Jordan's political inclination went far beyond and became
increasingly nationalistic and right-wing. These were of course not very good
prerequisites for resisting the temptation of the NS movement, in particular
since the conservative wing of the protestant church (to which he
adhered\footnote{The oldest son of the family patriarch Pascual Jorda was
brought up in the Lutheran faith of his foster mother, whereas all the other
children born within that marriage were raised in the Catholic faith.})
started to support Hitler in the 30's; in fact the behavior of both of the
traditional churches during the NS regime belongs to their darkest chapters.
Hitler presented his war of agression as a divine mission and considered
himself as an instrument of God's predestination (g\"{o}ttliche Vorsehung),
while almost all Christian churches were silent or even supportive.

Already in the late 20s Jordan published articles (under a pseudonym) of an
agressive and bellicose stance in journals dedicated to the spirit of German
Heritage; a characteristic ideology of right-wing people up to this day if one
looks at the present-day heritage foundations and their political power in the
US. It is unclear to what degree his more cosmopolitan academic peers in
G\"{o}ttingen knew about these activities. He considered the October
revolution and the founding of the Soviet Union as extremely worrisome
developments. One reason why Jordan succumbed to the NS-lure was perhaps the
idea that he could influence the new regime; his most bizarre project in this
direction was to convince the party leaders that modern physics as represented
by Einstein and especially the new Copenhagen brand of quantum theory was the
best antidote against the ``materialism of the Bolsheviks''. This explains
perhaps why he joined NS organisations at an early date when there was yet no
pressure to do so \cite{Wise}. He of course failed in his attempts; despite
verbal support\footnote{In contrast to Heisenberg he did not directly work on
any armament project but rather did most of his military service as a
meteorologist.} he gave to their nationalistic and bellicose propaganda and
even despite their very strong anti-communist and anti-Soviet stance with
which he fully agreed, the anti-semitism of the Nazis did not permit such a
viewpoint since they considered Einstein's relativity and the modern quantum
theory with its Copenhagen interpretations as incompatible with their
anti-semitic propaganda; one can also safely assume that the intense
collaboration with his Jewish colleagues made him appear less than trustworthy
in the eyes of the regime.

Jordan's career during the NS times ended practically in scientific isolation
at the small university of Rostock (his promotion to fill von Laue's position
in Berlin in 1944 was too late for a new start); he never received benefits
for his pro-NS convictions and the sympathy remained one-sided. Unlike the
mathematician Teichmueller, whose rabid anti-semitism led to the emptying of
the G\"{o}ttingen mathematics department, Jordan inflicted the damage mainly
on himself. The Nazis welcomed his verbal support, but he always remained a
somewhat suspicious character to them. As a result he was not called upon to
participate in war-related projects and spent most of those years in
scientific isolation. This is somewhat surprising in view of the fact that
Jordan, like nobody else, tried to convince the NS regime that fundamental
research should receive more support because of its potential weapons-related
applications; in these attempts he came closer to a ``star wars'' propagandist
of the Nazis than Heisenberg who headed the German uranium program but never
joined the party.

Jordan's party membership and his radical verbal support in several articles
got him into trouble after the war. For two years he was without any work and
even after his re-installment as a university professor he had to wait up to
1953 for the reinstatement of his full rights (e.g. to advise PhD candidates).
When his friend and colleague Wolfgang Pauli asked him after the war:
``Jordan, how could you write such things?'' Jordan retorted: ``Pauli, how
could you read such a thing?'' Without Heisenberg's and Pauli's help he would
not have been able to pass through the process of de-nazification (in the
jargon of those days Jordan got a ``Persilschein'', i.e. a whitewash paper)
and afterwards to be re-installed as a university professor. In Pauli's
acerbic way of dealing with such problems: ``Jordan is in the possession of a
pocket spectrometer by which he is able to distinguish intense brown from a
deep red''. ``Jordan served every regime trustfully'' is another of Pauli's
comments. Pauli recommended Jordan for a position at the University of Hamburg
and he also suggested that he should keep away from politics.

Jordan did not heed Pauli's advice for long; during the time of Konrad
Adenauer and the big debates about the re-armament of West Germany he became a
CDU member of parliament. His speech problem (he sometimes fell into a
stuttering mode which was quite painful for people who were not accustomed to
him) prevented him from becoming a scientific figurehead of the CDU party. At
that time of the re-armament issue there was a manifesto by the ``G\"{o}%
ttingen 18'' which was signed by all the famous names of the early days of the
university of G\"{o}ttingen quantum theory, including Max Born. Jordan
immediately wrote a counter article with the CDU parties blessing, in which he
severely criticized the 18 and claimed that by their action they endangered
world peace and stability. Max Born felt irritated by Jordan's article, but he
did not react in public against Jordan's opinion. What annoyed him especially
were Jordan's attempts to disclaim full responsibility for his article by
arguing that some of the misunderstandings resulted from the fact that it was
written in a hurry. But Born's wife Hedwig exposed her anger in a long letter
to Jordan in which she blamed him for ``deep misunderstanding of fundamental
issues''. She quoted excerpts from Jordan's books and wrote: ''Reines
Entsetzen packt mich, wenn ich in Ihren B\"{u}chern lese, wie da menschliches
Leid abgetan wird'' (pure horror overcomes me when I read in your books how
human suffering is taken lightly). Immediately after this episode she
collected all of Jordan's political writings and published them under the
title: ``Pascual Jordan, Propagandist im Sold der CDU'' (Pascual Jordan,
propagandist in the pay of the CDU) in the Deutsche Volkszeitung.

In the middle of the twenties the authors of the ``Dreimaennerarbeit'' were
proposed twice for the Nobel prize by Einstein, but understandably the support
for Jordan dwindled after the war. Nevertheless, in 1979 it was his former
colleage and meanwhile Nobel prize laureate Eugen Wigner who proposed him. But
at that time the Nobel committee was already considering second generation
candidates associated with the second phase of QFT which started after the war
with perturbative renormalization theory; there was hardly any topic left of
the first pioneering phase which was not already taken into account in
previous awards. Jordan did however receive several other honors, including
the Max-Planck-medal of the German Physical Society.

Jordan did not only have to cope with political satirical remarks such as
those from Pauli, but as a result of his neutrino theory of the photon
\cite{Neutrino} he also received some carnavalesque good-humored criticism as
in the following song (the melody is that of Mack the Knife) \cite{Pais}:

``Und Herr Jordan \ \ \ \ \ \ \ \ \ \ \ \ \ \ \ \ \ \ \ \ \ \ \ \ \ \ ``Mr. Jordan

Nimmt Neutrinos \ \ \ \ \ \ \ \ \ \ \ \ \ \ \ \ \ \ \ \ \ \ \ \ \ \ \ takes neutrinos

Und daraus baut \ \ \ \ \ \ \ \ \ \ \ \ \ \ \ \ \ \ \ \ \ \ \ \ \ \ \ \ and
from those he

Er das Licht
\ \ \ \ \ \ \ \ \ \ \ \ \ \ \ \ \ \ \ \ \ \ \ \ \ \ \ \ \ \ \ \ \ \ \ builds
the light.

Und sie fahren
\ \ \ \ \ \ \ \ \ \ \ \ \ \ \ \ \ \ \ \ \ \ \ \ \ \ \ \ \ \ \ \ And in pairs they

Stets in Paaren
\ \ \ \ \ \ \ \ \ \ \ \ \ \ \ \ \ \ \ \ \ \ \ \ \ \ \ \ \ \ \ always travel.

Ein Neutrino sieht man nicht.'' \ \ \ \ \ \ \ \ One neutrino's out of sight''.

Actually Jordan's idea is less ``crazy'' than it appears at first sight. In
the third section we will return to some interesting points concerning the
change of the concept of bound states in passing from quantum mechanics to
quantum field theory.

Although Jordan took (along with the majority of German physicists) a strong
position against those supporting the racist ``German Physics''\footnote{It
was Jordan's opinion that nationalistic and racist views had no place in
science; in his own bellicose style of ridicule (in this case especially
directed against nationalistic and racist stance of the mathematician
Biberbach): ``The differences among German and French mathematics are not any
more essential than the differences between German and French machine guns''.}
and in this way contributed to their downfall, he defended bellicose and
nationalistic positions and he certainly supported Hitler's war of aggression
against the ``Bolshevik peril''. The fact that he was a traditional religious
person and that several of the leading bishops in the protestant church were
pro Hitler had evidently a stronger effect on him than his friendship with his
Jewish colleagues, who by that time had mostly left Germany (in some cases he
tried to maintain a link through correspondence).

In contrast to Pauli who contributed to the second post war phase of QFT and
always followed the flow of ideas in QFT up to his early death, Jordan's
active participation in QFT stopped around the middle of the 30s and it seems
that he did not follow the development in that area. He turned his attention
to more mathematical and conceptual problems as well as to biology
\cite{Beyler} and psychology. His enduring interest in psychology was
presumably related to the psychological origins of his stuttering
handicap\footnote{One also should keep in mind that the interest in psychology
became a ``fashion'' among the Copenhagen physicists (notably Bohr and
Pauli).} which prevented him from using his elegant writing style in
discussions with his colleagues and communications with a wider audience; this
perhaps explains in part why even in the physics community his contributions
are not as well known as they deserve to be. In fact this handicap even
threatened his Habilitation (which was a necessary step for an academic
career) in G\"{o}ttingen. Jordan was informed by Franck (with whom he had
coauthored a book) that Niels Bohr had arranged a small amount of money for
Jordan which was to be used for getting some cure of his speech problem.
Wilhelm Lenz (whose assistant Jordan was for a short time after Pauli left)
suggested to go to the famous psychologist Adler. Jordan went to Vienna, but
we only know that he attended a lecture of Schr\"{o}dinger and criticized his
wave mechanics from the G\"{o}ttingen point of view; there is no record of
meetings with Adler.

His increasing withdrawal from the mainstream of quantum field theory and
particle physics in the 30s may have partially been the result of his
frustration that his influence on the NS regime was not what he had expected.
After the defeat of Germany in 1945 his attempts to account for his membership
in the Nazi party as well as the difficult task to make a living with the
weight of his past NS sympathies (which cost him his position as a university
professor for the first two years after the war) seriously impeded his
scientific activities.

Unlike the majority of the German population, for which the early Allied
re-education effort (which was abandoned after a few years) to rid society of
aggressive militaristic and racist ideas was a huge success so that the
subsequent change of US policy in favor of re-armament of West Germany ran
into serious opposition during the Adenauer period, Jordan did not completely
abandon his militaristic and rightwing outlook. In the 50s he joined the CDU
party where he had to undergo the least amount of change \footnote{The
leadership of the CDU recently supported Bush's war in Iraq (against the
majority of its voters).}, thus forgetting Pauli's admonitions in favor of
political abstinence.

All the protagonists of those pioneering days of quantum physics have been
commemorated in centennials except Pascual Jordan who, as the result of the
history we have described, apparantly remained a somewhat ``sticky'' problem
despite Pauli's intercession by stating ``it would be incorrect for West
Germany to ignore a person like P. Jordan''. His postwar scientific activities
consisted mainly in creating and arranging material support (by grants from
Academies and Industry as well as from the US Airforce) for a very successful
group of highly motivated and talented young researchers in the area of
General Relativity who became internationally known (Engelbert Schuecking,
Juergen Ehlers,..) and attracted famous visitors especially from Peter
Bergmann's group (Rainer Sachs,....). In this indirect way there is a
connection between Jordan's post war activities in general relativity and the
new Albert Einstein Institute in Golm (Potsdam). This somewhat meandering path
leads from Jordan's Seminar in Hamburg through universities in Texas (where
most of its members got positions), and then via the astrophysics in Garching
(where Ehlers took up a position in 1971) to the AEI for Gravitational Physics
of the MPI where Ehlers became the founding director in 1995.

Jordan died in 1980 (while working on his pet theory of gravitation with a
time-dependent gravitational coupling); his post war work never reached the
level of the papers from those glorious years 1925-1930 or his subsequent
rather deep pre-war mathematical physics contributions. In the words of Silvan
Schweber in his history of quantum electrodynamics, Jordan became the ``unsung
hero'' of a glorious epoch of physics which led to the demise of one of its
main architects.

It is however fair to note that with the exception of Max Born, Jordan's other
collaborators, especially von Neumann and Wigner, shared the bellicose kind of
anti-communism; Wigner later became an ardent defendor of the Vietnam war.
Since both of them came from a cosmopolitan Jewish family background, their
anti-communist fervor probably had its roots in their experience with the
radical post World War I Bela Kuhn regime in the Hungarian part of the
decaying Habsburg empire. Hence Jordan's right wing anti-communist views posed
no friction during the time of his collaboration with Wigner and von Neumann.

The cultural and scientific achievements in a destroyed and humiliated Germany
of the post World War I Weimar republic within a short period of 15 years
belong to the more impressive parts of mankind's evolution and Jordan, despite
his nationalistic political viewpoints is nevertheless part of that heritage.

\section{Jordan's contributions to Quantum Mechanics and to the first phase of
Quantum Field Theory}

In the following I will give a more detailed account of the content of some of
Jordan's most seminal contributions to QM and QFT. Like the problem of
appreciating the significance of the conceptual step from Lorentz to Einstein
who share the same transformation formula, there are several instances where
important progress is not primarily reflected in formulae but in paradigmatic
changes of interpretations. Since the fashions of the last decades have led to
an atrophy\ of the art of interpretation as compared to calculation, a careful
presentation of the conceptual progress achieved in Jordan's work may be helpful.

The situation which Jordan was confronting, after he was called upon by Born
to collaborate on the mathematical and conceptual underpinnings of
Heisenberg's pivotal work, was as follows. Heisenberg had gone far beyond the
somewhat vague correspondence principle by his proposal to substitute for the
unobservable classical particle variables q,p a novel kind of in principle
observable set of complex amplitudes obeying a non-commutative multiplication
law. These quantities were supposed to satisfy a ``quantum condition'' which
formally resembles the Thomas-Kuhn sum rule\footnote{In comparision with
Lorentz versus Einstein where the significant conceptual difference did not
necessitate a different notation for the transformation formula, the new
Heisenberg world required a totally different notation as that in the papers
of Thomas and Kuhn.} for harmonic oscillators but which, in terms of the new
quantities was to be regarded as a basic universal quantum law. Born
immediately identified the new quantities as matrices. Moreover he found
Heisenberg's quantum condition to assert that all diagonal elements of $pq-qp$
are equal to $\frac{\hbar}{i},$ whereupon he conjectured the famous
commutation relation.

Jordan quickly succeeded to prove the vanishing of the off-diagonal elements
as a consequence of the equations of motion and an ingenious algebraic
argument \cite{B-J}. From that time on the (matrix-form of the) commutation
relation was the new principle of quantum mechanics. The subsequent
``Dreim\"{a}nnerarbeit'' \cite{B-H-J} extended the setting of quantum
mechanics to systems with many degrees of freedom.

Shortly after this paper, Jordan together with Heisenberg \cite{H-J}
demonstrated the power of the new formalism by presenting the first fully
quantum-mechanical treatment of the anomalous Zeeman effect based on the new
electron spin hypothesis of Goudsmit and Uhlenbeck. In this article the
authors left no doubt that they considered this as temporary working
hypothesis short of a relativistic description of the electrons intrinsic
angular momentum; a problem which still had to wait three years before Dirac
finally laid it to rest.

Most of Jordan's works after 1926 deal with ``quantization of wave fields''
i.e. with quantum field theory. There is however one important later
contribution which addresses the foundations of quantum physics and led to an
algebraic structure which bears his name, the so-called Jordan algebras. If
one accepts von Neumann's axiomatic framework of quantum theory, which
identifies observables with Hermitian operators acting in a Hilbert space, one
would like to have at one's command a multiplication law which converts two
observables into a composed observable. This is achieved by taking the
anti-commutator and Jordan posed the problem of unravelling the algebraic
structure which one obtains if one disposes of the Hilbert space setting and
axiomatizes this abstract algebraic structure. The result was a commutative
but not associative new structure, the so-called Jordan algebras \cite{J}.
This attracted the interest of von Neumann and Wigner and led to some profound
mathematical results. In a joint paper \cite{J-N-W} they proved that (apart
from a very special exceptional case that for finite dimensional Jordan
algebras without loss of generality one obtains ordinary matrix algebras with
the anti-commutator composition law. This showed that the standard quantum
theory setting of operator algebras in Hilbert space as axiomatized by von
Neumann was more natural and stable against modifications than one had reasons
to expect on the basis of Heisenberg's observable credo. This work was later
extended to the infinite dimensional realm (by making suitable topological
assumptions about Jordan algebras) without encountering any additional
exception \cite{Alfsen}. 

Works on Jordan algebras, to the extent that they are physically motivated,
should be viewed in line with other attempts to obtain a better understanding
of the superposition principle\footnote{The total Hilbert space of a quantum
physical system decomposes into coherent Hilbert subspaces which are subject
to the superposition principle (i.e. the von Neumann axiom that with two
physically realizable vectors any vector in their linear span is also
physically realizable) is valid.}. The latter is the fact that states in the
sense of positive linear forms on algebras permit an (in general nonunique)
interpretation as expectation values in vector states (or in terms of density
matrices in case the states are impure) in a Hilbert space which carries an
operator algebra realizations of the abstract algebra such that the linear
combination of two vectors defines again a (pure) state. Since quantum states
are conceptually very different from classical linear waves, the history of
quantum physics up to this date is rich of attempts which try to make the
quantum superposition principle physically more palatable. The dual
algebra-state relation permits to explore this problem either on the algebraic
side in the spirit of Jordan or on the side of the structure of the state
space \cite{Bel}. The farthest going results on the side of states was
obtained by A. Connes \cite{Connes} who succeeded to characterize operator
(von Neumann) algebras in terms of the facial substructure of their convex
state spaces (the predual of algebra). \ 

In the remainder of this section we recall Jordan's most important and
enduring discovery namely that of quantum field theory. The conceptual
difference of what Jordan did as compared with Dirac's ``second quantization''
is somewhat subtle. The latter was an artful transcription (the Fock space
formalism was not yet available) of the Schr\"{o}dinger multi-particle setting
into a form of what in modern terminology is called the occupation number
representation; for the formal backup of this step Dirac used his version of
the transformation theory. 

Jordan had the bold idea to view the one-particle Schr\"{o}dinger wave
function as a classical wave equation to be subjected to the extended
quantization rules which one would naturally impose on the canonical field
formalism derived from classical Lagrangians. By a detailed calculation Jordan
\cite{Wellen} showed that this procedure, projected onto the n-particle
subspace, is equivalent to Dirac's occupation number description. At that time
Jordan met Oscar Klein in Copenhagen, both of them were guests of Niels Bohr.
In a widely acclaimed paper they jointly extended the field quantization
formalism to the case with interactions \cite{J-K}. 

Jordan considered it as a significant advantage that his new viewpoint allowed
to incorporate ``Pauli statistics'' together with Bose-Einstein statistics
into one and the same formalism; in fact some of the wave field quantization
ideas were already contained in the concluding sections of \cite{Gas} which
delt with anti-commutation relations. He returned to this subject in a joint
work with Wigner (submitted in Januaray 1928) which contains significant
extensions and clarifications \cite{J-W}. This paper is not only quoted as an
alternative approach to Dirac's presentation of anti-commutation relations,
the Jordan-Wigner method also received particular attention in connection with
nonlocal transformations which are capable of changing commutation relations.
In this paper Jordan and Wigner discovered an abstract (but highly ambiguous)
method to write Fermions in terms of ''Paulions'' i.e Pauli matrices (which
however as a result of Jordan's lifelong love for quaternions appear in a
quaternionic camouflage). The ordering prescription which they need in order
to write concrete Fermion formulas in the Hilbert space of a discrete array of
Pauli spin matrices becomes physically unique in case of the presence of a
natural spatial ordering as in the transfer matrix formalism of the 2-dim.
Lenz-Ising model \cite{Lieb}. This kind of nonlocal formula involving a line
integral (a sum in the case of a lattice model) became the prototype of
statistics changing transformations (bosonization/fermionization) in d=!+1
models of quantum field theory \cite{Thirring}\cite{Rothe} and condensed
matter physics \cite{Lut}

It is interesting to observe Dirac's reaction; different from the general
enthusiastic acceptance of Jordan's approach he was somewhat disappointed that
Jordan's version of second quantization\footnote{The present day usage of this
terminology is in the sense of Jordan's field quantization approach.} did not
yield more results beyond the ones he himself had worked out. He also was not
much impressed by the formal incorporation of the anti-commutation structure
into Jordan's wave field quantization setting. His first complaint was of
cause rendered unjustified as soon as the new field quantization approach was
applied to the relativistic setting \cite{P-J} were the interaction-caused
vacuum polarization and real particle creation left no alternative. I think
that his second point was more of an estetical and philosophical nature since
it appears somewhat unnatural to invent a fictitious classical reality (namely
classical Fermions, nowadays often equipped with a Grassmann structure) just
in order to be able to grind halfinteger spin through the same quantization
mill. As will be explained in the next section the new approach of Local
Quantum Physics \cite{Haag} avoids this problem by its dichotomy of local
observables and local field systems as irreducible representation
(superselection) sectors of these observables.

The continuation of the field quantization saga is well known. After the 1928
Jordan-Pauli paper on the spacetime treatment (overcoming the canonical
formalism -aused separation into space and time) in which the free field
Jordan-Pauli commutator functions appeared for the first time, the field
quantization torch was taken over by Heisenberg and Pauli. The subsequent
study of properties of the vacuum problems in QFT brought the QFT train to a
grinding halt. After the second world war the new locomotive of
renormalization led to a remarkable recovery and a new faith in the underlying
principles of QFT but the team running the train had changed. Instead of the
meanwhile grown up members of the Knabenphysik of the Goettingen days, the
physicists in the driving seat consisted to a large degree of young Americans;
the physics language changed from German to English.

In 1929 at a conference in Kharkov\footnote{Landau, after his return from
Copenhagen, went to the university of Kharkov which for a short time became
the ``Mecca'' of particle physics in the USSR.}, Jordan gave a remarkable
plenary talk \cite{Kharkov} (the conference language at that time was still
German). In a way it marks the culmination of the first pioneering phase of
QFT; but it also already raised some of the questions which were partially
answered almost 20 years later in the second phase of development (i.e.
renormalized perturbation theory, gauge theory). In his talk Jordan reviews in
a very profound and at the same time simple fashion the revolutionary steps
from the days of matrix mechanics to the subsequent formulation of
basis-independent abstract operators (the transformation theory which he
shares with Dirac) and steers then right into the presentation of the most
important and characteristic of all properties which set QFT apart from QM:
Locality and Causality as well as the inexorably related Vacuum Polarization.
Already one year before in his Habilitationsschrift he identified the two
aspects of relativistic causality namely the statistical independence for
spacelike separations (Einstein causality, commutance of observables) and the
complete determination in timelike directions (the causal shadow property) as
playing a crucial role in the new quantum field theory. He ends his
presentation by emphasizing that even with all the progress already achieved
and that expected to clarify some remaining unsatisfactory features of gauge
invariance (\textit{Die noch bestehenden Unvollkommenheiten, betreffs
Eichinvarianz, Integrationstechnik usw., duerften bald erledigt sein)}, one
still has to confront the following problem\footnote{Here we have actualized
in brackets the content of this sentence since ``QED'' (the only existing QFT
in those days) was used in the same way as ``QFT'' in present days.}:
\textit{Man wird wohl in Zukunft den Aufbau in zwei getrennten Schritten ganz
vermeiden muessen, und in einem Zuge, ohne klassisch-korrespondenzmaessige
Kruecken, eine reine Quantentheorie der Elektrizitaet zu formulieren
versuchen. Aber das ist Zukunftsmusik.} (In the future one perhaps will have
to avoid the construction in two separated steps and rather have to approach
the problem of formulating a pure quantum theory of electricity (a pure
quantum field theory) in one swoop, without the crutches of classical
correspondences. But this is part of a future tune.)

He returns on this point several times, using slightly different formulations
(....\textit{muss aus sich selbst heraus neue Wege finden) }for a plea towards
a future intrinsic formulation of QFT which does not have to take recourse to
quantization which requires starting with a classical analog.

These statements are even more remarkable if one realizes that they come from
the protagonist of field quantization only two years after this pivotal
discovery. When I accidentally came across the written account of this Kharkov
talk, I was almost as surprised as I was many years before when it became
known that Oscar Klein (with whom Jordan collaborated in the 30s in
Copenhagen) had very advanced ideas about nonabelian gauge theory (which were
published in the proceedings of the 1939 Warsaw international conference
\cite{Klein}). Apparently all of the classical aspects of nonabelian gauge
theories and some quantum aspects were known before the second world war.
Apart from QED the postwar interest developed away from gauge theories into
pion-nuclear physics. The great era of gauge theory in connection with strong
interactions had to wait 3 decades, and by that time Klein's work was
completely forgotten and played no role in the birth of QCD.

In the following two section we will try to convince the reader that Jordan's
expectations about a radically different realization of the underlying
principles of QFT have meanwhile been realized in Local Quantum Physics (LQP)
\cite{Haag}. This approach has rejuvenated QFT by leading to a wealth of new
questions which are presently being investigated by new methods. Again there
is no direct historical connection between Jordan's ideas and modern LQP.
Rather these episodes confirm the belief that although important ideas in the
exact sciences may get lost in certain situations, sooner or later they will
be rediscovered and expanded.

Jordan's expectation about a rapid understanding of the ``imperfections of
gauge theories'' at the time of his Kharkov talk may have been a bit
optimistic since the Gupta-Bleuler formalism only appeared 20 years later
\cite{GP}. For Jordan gauge theory was an important issue already in 1929.

Finally one should add one more topic which shows that Jordan and Dirac had a
very similar taste for what both considered the important problems of the
times (in addition to: statistics and (anti)commutation relations, operator
transformation theory, second quantization): magnetic monopoles. In this case
Dirac is the clear protagonist of this idea, but Jordan found an interesting
very different algebraic argument for the monopole quantization which he based
on the algebraic structure of bilinear gauge invariants \cite{Mono} (in the
setting in which Mandelstam 30 years later attempted a gauge invariant
formulation) adapted to the magnetic flux through a tetrahedron. The problem
of how these kind of arguments have to be amended in order to take account of
renormalization has according to my best knowledge not been satisfactorily answered.

Since Jordan left the area of QFT in the middle of the 30s and became
disconnected from the discoveries thereafter, we do not know how he would have
reacted to the amazing progress after the war brought about by renormalization
theory. But contrary to expectations of the leading theorist during the 30s
this progress was obtained by a careful conceptual distinction between formal
and physical (observed) Lagrangian parameters and a systematic extension of
the existing formalism i.e. it was achieved in a rather conservative manner
\cite{Wein}. Renormalization theory is too conservative as far as field
quantization is concerned in order to serve as a candidate which could fit
Jordan's.radical expectations. In strange contrast to his political stance, in
physics Jordan was a visionary revolutionary.

\section{Jordan's quantum field theoretical legacy in past and present achievements}

In times of stagnation and crisis as the one we presently face in the post
standard model era of particle physics, it is helpful to look back at how the
protagonists of quantum field theory viewed the future and what became of
their ideas and expectations. Perhaps the past, if looked upon with care and
hindsight, may teach us where we possibly took a wrong turn and what
alternative path was available.

Jordan's Kharkov talk marks a high point in presentations of developments of
QFT and in particular his own participation in the shaping of physical
concepts, but in a way it also can be viewed as rounding off the pioneering
stage of QFT; in the nearly 20 years up to the beginnings of the second stage
of perturbative renormalization theory due to Feynmann, Schwinger and Dyson
(with important conceptual methodical and computational contributions by
Kramers, Tomonaga, Bethe and many others), there was a kind of conceptional
lull apart from some isolated but important contributions.

One significant exception whose full potential was appreciated only much later
is Wigner's famous 1939 group representation-theoretical approach
\cite{Wigner} to particles and their classification.

In fact Wigner accomplished in a very limited context to obtain the kind of
intrinsic formulation that Jordan in his Kharkov talk was contemplating for a
post quantized wave field approach. Instead of describing particles in terms
of wave equations, which leads to many different-looking but nevertheless
physically equivalent formulations, Wigner showed that one obtains a
completely intrinsic and unique description of the possible wave function of
relativistic particles without the necessity of quantizing classical
structures by simply classifying irreducible positive energy ray
representations of the Poincar\'{e} group\footnote{Due to the projective
nature of quantum mechanical states, one has to classify projective (ray)
representations, but for many groups including the Poincar\'{e} group
(depending on group cohomology) this can be encoded into vector
representations of their universal covering.}. One first determines the
irreducible representations of the (universal covering of the) connected part
and then extends to the full Poincar\'{e} group with the help of the parity
and (anti-unitary) time reversal transformations which in some cases requires
a doubling of the representation space. The standard way to come from the
unique Wigner (m,s) representations to quantized free fields is well known
\cite{Wein}. The important step is the classification of u- and v-
intertwiners which relate the unique $(m,s)$ Wigner representation to
covariant (spinor,tensor) representations; there is a countably infinite
family of covariant representations originating from one $(m,s)$ Wigner
representation. The infinitely many pointlike fields are all singular
operators (operator-valued distributions) in the same Hilbert space, in fact
they turn out to be different ``field coordinatizations'' of a net of local
algebras which is uniquely associated with the Wigner representation. However
they do not exhaust the possibilities of pontlike coordinatizations of
algebras but only constitute a linear subset; nonlinear coordinatizations are
obtained by taking Wick polynomials in those fields.

The intrinsic field-coordinatization-independent description in the spirit of
Jordan's Kharkov talk is that of the spacetime-indexed net of local
algebras\footnote{The impression that Haag was led to his intrinsic
formulation as a result of Jordan's 1929 Kharkov remarks is of course wrong,
inasmuch as Yang and Mills were unaware of the content of Klein's 1939 Warsaw
talk on nonabelian gauge fields. But it shows that good ideas often have
prophetic precursors.} \cite{Haag}. If we had to construct first pointlike
fields in order to find the net of algebras associated with the (m,s) Wigner
representation, we would not have gained much as compared to the standard
approach in \cite{Wein}. There exists however a direct path from the unique
$(m,s)$ Wigner representation to the unique net of operator algebras via
modular theory \cite{BGL}\cite{F-S1} without passing through nonunique field coordinatizations.

The idea is the following. The first step consists in characterizing a real
Hilbert subspace $H_{r}(W)\subset H_{Wig}$ of the complex Wigner space which
describes wave functions which are modular localized in a (Rindler) wedge
region $x\geq\left|  t\right|  ;y,z$ arbitrary. This is done with the help of
an involutive closed antilinear Tomita $S_{T}$-operator (the +1 eigenstates of
$S_{T}$) which is defined in terms of the Wigner representation theory (see
next section). Being in the domain of $S_{T}$, the wedge-localized momentum
space Wigner wave functions possess analyticity properties which permit a
continuation to negative energies within the complex mass shell. This is
reminiscent of the Klein paradox, which links positive with negative energies
and drives the situation away from a standard quantum mechanical one-particle
setting, except that in our case there is no x-space wave function and no
coupling to external fields. The analytic continuation to negative energies in
the $H_{r}(W)$ wave functions resembles a crossing operation \cite{F-S1}. In
fact it turns out that these modular localization properties preempt among
other things the spin-statistics connection and the TCP theorem which one
usually derives in the full quantum field theory.

The remaining step in the construction of the net of operator algebras is
quite simple. For integer spin the Weyl functor maps the wedge-localized real
subspace into the wedge-localized operator (von Neumann) subalgebra
$\mathcal{A}(W)$ of the algebra of all bounded operators $B(H)$ in Fock space,
and the CAR functor accomplishes the analogous result for halfinteger
spin\footnote{These functors map the category of real Hilbert spaces into von
Neumann operator algebras in such a way that spatial inclusions pass into
algebraic ones. It only exists in the absence of interactions and often
refered to as ``second quantization''. In the words of E. Nelson: (first)
quantization is a mystery, but second quantization is a functor.}. The full
net is obtained from the net of wedge-localized algebras by algebraic
intersection. It is customary to consider the causally closed
Lorentz-covariant family of double cone (diamond-shaped) regions $\mathcal{O}$
as the generating sets of Minkowski spacetime in which case the von Neumann
algebras $\mathcal{A(O)}$ are the smallest building blocks of the net. The
localization-core of double cones of arbitrary size is what one gets if one
scales down the size to to zero, namely a spacetime point.

For the standard Wigner particle representations with s=(half)integer (in case
of m=0, s is the helicity) this modular construction is mainly of
methodological interest since the covariantization approach \cite{Wein} leads
to free fields which also provide a complete physical description. The main
advantage of the operator algebra approach is that the somewhat singular
nature (which unleashes its full ``nastiness'' only if one uses these singular
objects for implementing interactions) of pointlike fields are avoided. The
intersections of operator algebras whose localization regions intersect in a
point consists of the trivial algebra (multiples of the identity) and hence
pointlike fields are idealized singular objects associated to the algebras in
the sense that after smearing with $\mathcal{O}$-supported test functions one
obtains unbounded operators affiliated with $\mathcal{A(O).}$ This is what is
meant by saying that pointlike fields constitute a ``singular
coordinatization'' of the net of operator algebras. Jordan did of course
obtain his quantum wave fields not as a singular coordinatization of algebras
but by quantizing classical wave fields. The singular nature of these
quantized fields (later mathematical research identified them as
operator-valued distributions) may have been a motivation for contemplating an
implementation of the physical principles underlying QFT without quantization ``crutches''.

The crucial question is whether such a viewpoint also exists in the presence
of interactions. Before we address this problem (which will be the central
issue of this and the last section), it is important to note that there are
two special positive energy representations which do not belong to the above
standard cases. One of these special cases is Wigner's famous zero mass
``continuous spin'' representation which has an internal structure described
by a helicity spectrum which resembles those infinite towers which
characterize the ``stringyness'' of string theory. The application of the
modular localization theory reveals that their Wigner spaces \textit{do not
admit nontrivial compactly localized subspaces} $H_{r}(\mathcal{O})$ for e.g.
$\mathcal{O}=$ double cone; rather their smallest nontrivial localization
regions are (arbitrarily narrow) \textit{spacelike cones}\footnote{A spacelike
cone is a noncompact causally complete region which consists of a conic
neighbourhood of a semiinfinite spacelike line (the core) with the line
starting at the apex of the cone.} (i.e. $O=$ spacelike cone) \textit{whose
localization-cores are semiinfinite (straight) open ``strings''}
\cite{BGL}\cite{F-S1}. Although these ``Wigner strings'' have a vague
resemblence to the classical Nambu-Goto string (the ur-version of string
theory) in that they are objects which possess a rich internal spacetime
related structure, the differences between them are much more important in the
present context. \textit{A Wigner string is a pure quantum string}, i.e. it
can not be interpreted as the result of a quantization applied to a classical
object and hence \textit{does not permit a Lagrangian description}. This is
the reason why in Weinberg's book \cite{Wein} (whose main theme in the first
section is the use of Wigner's theory as an additional support for Lagrangian
quantization) the zero mass helicity tower representations play no role. The
corresponding quantum fields are (as in the standard case) objects which are
sharply localized on the localization core of the spacelike cone regions and
have to be smeared in the localization-core in order to obtain Fock space
operators whose one-fold application to the vacuum generate the localized
Wigner subspaces. The standard $u,v$ intertwiner construction \cite{Wein} from
Wigner- to covariant- representations has to be generalized in a nontrivial
manner \cite{M-S}. If Jordan would have known these consequences of the later
theory of his colleague Wigner, he could have noticed that his farsighted
post-quantization outlook would have been supported by this example which
(aside from the avoidance of singular coordinatizations) does not admit a
quantization description.

The other special case is that of massive particles in d=1+2; it does not
appear in Wigner's d=1+3 list but was later analyzed by Bargman \cite{Barg}.
It turns out that in this case the localization-core of the smallest
nontrivial localization region is also semiinfinite stringlike, but in this
case the mechanism is not related to the rich internal structure but rather
results from the fact that the d=1+2 Poincar\'{e} group has an infinite
sheeted universal covering \cite{Mund3}. The projective nature of quantum
theory always requires the use of the covering in order to include with the
generic case. The latter leads to (m,s) representations with s=real,
unquantized. For s=halfinteger the ``anyonic'' spin reduces to Bosons/Fermion
Wigner spaces which admit nontrivial compact localization (with pointlike
localization-cores). 

The anyonic strings for $s\neq\,$halfinteger are quite different from those of
the helicity towers; they are more like those strings which Mandelstam
envisaged for the description of charge-carrying objects in gauge theories
except that their ``living space'' requires the topology of an auxiliary
2-dimensional de Sitter space in order to characterize the asymptotic string
directions relative to a reference direction \cite{Mund3}. In contrast to the
tensor product structure of the ``string field theory'' of the Wigner strings,
the repeated applications of \ ``free'' anyonic string operators would not be
compatible with tensor product factorization, i.e. they are not free fields in
the usual technical sense. This is related to the fact that they obey braid
group commutation relations which inevitably cause vacuum polarization clouds
in their one-field states obtained from their application to the
vacuum\footnote{Braid group statistics can only be maintained in the presence
of vacuum polarization clouds, even if there are no genuine interactions (zero
cross sections), i.e. even in the case of ``free anyons'' \cite{Mund2}.}; in
fact their nonrelativistic limit would not allow a quantum-mechanical
description if one insists upholding (as in case of Fermions/Bosons) the
spin-statistics connection. The interesting question in the absence of a
quantization procedure is of course whether one can generate the associated
operator algebras by ``smearing'' singular field coordinates with covariance
properties which reflect their stringlike nature. In the next section we will
present powerful ideas from modular theory which indicate a solution for this
kind of problem.

There are two important messages coming from the Wigner theory which
illustrate the trans-quantization idea of Jordan in his plea for abandoning
the quantization ``crutches''. The first message is that the non-unique
singular pointlike free fields can be avoided in favor of a direct
construction of the unique local net of algebras which they generate. The
second message is that \textit{the most general localization cores (associated
to optimally localized subspaces of Wigner positive energy representations)
are semiinfinite stringlike }and in this case these objects do not possess a
classical counterpart. This is remarkable since this stringlike geometry is
not there because some physicists set out to study what happens if one passes
from point- to string-like\textit{ }extensions, but rather because the
stability principle underlying the positive energy representations permit only
point- and string-like realizations! Whereas in the stringlike cases the
double cone localized one-particle subspaces are trivial (the nullvector), one
expects that the associated field theory also admits double-cone localized
observables (string/anti-strings), so that the string localization refers to
charge-carrying fields and the neutral local observables remain compactly
localizable with pointlike localization-cores. The key question concerning
local quantum physics is therefore: is semiinfinite stringlike localization
the worst localization-core which can happen for charge-carrying optimally
localized objects in LQP? Under somewhat more restrictive assumption (positive
energy with mass gaps) the answer is affirmative; this is a fundamental result
of Buchholz and Fredenhagen \cite{Haag}.

The generalization of this operator algebraic approach to the case of
interaction has its beginnings in the late 50s when Haag formulated the idea
that relativistic causality provides a very rich local substructure to the
global operator algebra \cite{H57}. Other operator algebraic approaches (e.g.
that of Irving Segal) failed to appreciate the significance of this local
algebraic substructure . In the beginning of the 60s the first systematic
accounts of these ideas appeared \cite{H-S}. \ The local algebras had some
very unfamiliar mathematical properties which were unknown in quantum
mechanics and are (like all properties which distinguish QFT from QM) related
to the physical phenomenon of vacuum polarization resulting from causality and
leading to type III von Neumann algebras \cite{Araki}.

By the middle 60s the main mathematical and conceptual ingredients were in
place (Haag-Kastler, Borchers). The new credo of local quantum physics as the
intrinsic approach without quantization crutches was that the structural
consequences of the physical principles of QFT are most naturally derived by a
dichotomy between local observables and charge-carrying fields. In the setting
of local nets this amounts to an inclusion of algebraic nets. The smaller net
consists of local observables which fulfill obvious causality requirements
which constitute the only firm link with classical physics\footnote{Instead of
quantizing classical field theories, one adapts the causality principles to
the setting of LQP and views the problem of classifying and constructing
models as one which has to be handled inside LQP. Since the majority of
particle physicists have a fair knowledge about chiral conformal theories, it
may be helpful to point out that in many respects LQP may be viewed as a
generalization of the algebraic ideas (representing algebras instead of
quantizing Lagrangians) used in chiral theories.}. Here the physical intuition
about causality and stability in local quantum physics enters; however the
somewhat mysterious concept of inner symmetries (additive charge quantum
numbers and their isospin like generalizations) which attributes charge
quantum numbers to particles and fields is not used because one of the reasons
for this dichotomic approach is precisely to explain field and particle
statistics and inner symmetries as a manifestation of spacetime-organized
quantum matter. 

The second step in the observable-field dichotomy is (in Wigner's
representation theoretical spirit) the classification of all causally
localizable representations of this observable net; for zero temperature
particle physics this amounts to finding all positive energy representations
of the observable net which are locally unitarily equivalent to the vacuum
representation. 

The last step is a mathematical conceptual one: organize these representations
in such a way that they become the superselection sectors of the
representation of a larger algebra, the so-called field algebra\footnote{Here
the meaning of ``field'' is not in the pointlike singular sense, but rather
stands for charge-carrying operators as opposed to the neutral operators which
generate the observable algebra.}. This step is achieved by introducing charge
carrying intertwiners between the superselected (not coherently superposable)
representations. There is some liberty in the choice of these intertwiners
which leads to different relative spacelike commutation structures among
fields which carry different charges; this can be settled by natural
conventions (achievable by so-called Klein transformations). The end result is
very beautiful; one obtains a field algebra i.e. a spacetime indexed net which
has no further inequivalent localizable representations i.e. the dichotomy
allows no further extension \cite{D-R}. The increase of structural
understanding is enormous. 

On the one hand one rederived properties, as the connection between spin and
statistics and the TCP theorem, which already were rigorously obtained in the
general pointlike setting of Wightman, but now without making any assumptions
about non-observable fields\footnote{In Pauli's derivation which was later
refined by Jost \cite{S-W}, one had to assume that the charge-carrying fields
obey either spacelike commutation- or anticommutation-relations.}. 

On the other hand one learns that the concept of inner symmetry in particle
physics is preempted in the net structure of observables; the specific
symmetry group for a given observable net is a compact subgroup of some SU(n)
for large enough n which can be computed solely from the observable data
\cite{D-R}. There is no better verbal way to appreciate this intellectual
achievement than by quoting Marc Kac's aphorism about the mathematical physics
aspects of acoustics: ``How to hear the shape of a drum?'' The D-R
construction is like a reconstruction of the full field theoretic data from
their observable shadow.

It is very revealing to compare this de-mystification of internal symmetry
with the problems at the time its beginning in the middle 30s. Most particle
physicists learned from textbooks that at that time Heisenberg had the idea
that collecting the proton and neutron into a doublet facilitate the
understanding of nuclear forces. What is not generally known is that even
before Heisenberg wrote his second paper in which he added the group theoretic
formalization of isospin, Jordan investigated the relation between the
representations of the permutation group and internal SU(n) symmetry
(multiplet structure) of his quantized wave field formalism in a remarkable
paper \cite{Sym}. Since the mathematically controllable part of the Lagrangian
field formalism was restricted to bilinears (free fields), his point of
departure was the algebraic structure of bilinears in momentum-space creation
and annihilation operators. In this way he obtained the best possible field
theoretic restrictions on the general connections between the representation
theory of the permutation group (statistics) and the internal symmetry groups
(the multiplicity structure of fields) in a Lagrangian setting. Some time
after Heisenberg's SU(2) isospin paper, Schwinger rediscovered (without
reference to Jordan) the SU(n) results using a similar method. Schwinger's
contribution became widely known (especially through a book written by
Biedenharn and van Dam \cite{Schwinger}).

A complete de-mystification of the internal symmetry issue was not possible at
the time of the Jordan's and Schwinger's contributions since the
interrelations between causality and localization properties with a
generalized concept of charges through the omnipotent vacuum polarization
properties were not really understood. Even though in the 1930s most of the
physical principles of QFT were in place, several important concepts which
reveal deep connections between them were still missing; the complete
de-mystification of interal symmetry in terms of the causal observable
strucure had to wait more than 50 years \cite{D-R}.

Another important corner-stone was added to particle physics when Wheeler and
Heisenberg introduced the scattering matrix. The S-matrix, as it was later
referred to, was not only an invariant measure of interactions between
particles, but became the dominant observable of particle physics in terms of
which most experimental results could be analyzed. Its formal use in
renormalized perturbation theory by Dyson contributed a significant amount of
conceptual order to the renormalization recipes. Its detailed relation to
quantum fields turned out to be one of particle physics profound problems.

A complete solution was attained only when the asymptotic convergence in the
LSZ theory was finally backed up by mathematical proofs which showed that the
timelike asymptotic convergence and the invariance of the asymptotic
transition operator S is a consequence of spacelike causality and suitable
spectrum conditions (existence of a mass gap). For the first time conceptual
clarity in the particle-field relation substituted the intuitive but somewhat
ambiguous particle-wave duality.

Scattering theory provided a Fock space structure to the representation space
of a QFT even in the presence of interactions; if the application of operators
to the vacuum state leads to vectors which have a one-particle Wigner
component, then the locality structure of QFT guarantees the presence of
(anti)symmetrized tensor product states of Wigner particles. This is more
specific and precise than Jordan's ``corpuscles'' from quantized waves. There
is a fine point here which tends to be sometimes overlooked even in rather
sophisticated contemporary work. The multi-particle interpretation of state
vectors is limited to Lorentz frames and there is no reason to believe that
e.g. an LSZ asymptotic behavior holds in the time in the Rindler world of
uniformly accelerated Unruh observers \cite{Unruh}. As a result the particle
terminology in those cases does not seem to be appropriate; the arguments are
based on linear (free) systems whose Rindler excitations are identified with
particles\footnote{Therefore it may be somewhat misleading to use the
terminology of ``particles'' in connection with the Rindler-Unruh situation
(as one find in the literature \cite{Wald}\cite{Halvor1}) since there is no
reason to believe that the Hilbert space analyzed in terms of thermal Unruh
excitations has a Fock space structure (apart from the atypical case of
non-interacting quantum matter).} but one cannot expect an LSZ mutiparticle
structure to hold in the frame of an uniformly accelerated Unruh observer (or
in general curved spacetime situations).

The omnipresence of vacuum polarization renders any quantum mechanical
bound-state picture obsolete; what gets fused are (superselected) generalized
charges and not particles. Whereas the relation among particles is basically
``democratic'' (``nuclear democracy''), that between charges is hierarchical.
The only hierarchy among particles arises from charges which they carry; it
has nothing to do with the quantum mechanical distinction between elementary
versus bound which is based on particle number conservation. 

In the Lagrangian approach a hierarchy between a ``would be'' fundamental
field and its fused composites has been introduced via the quantization
procedure; but even in such a context it is not clear whether this distinction
has an intrinsic physical meaning; among the many possible scenarios it could
happen that the objects carrying the fundamental charges are outside the
Lagrangian description. This happens e.g. with the Sine-Gordon Lagrangian
where the Lagrangian field creates a local net of operator algebras which in
addition to its vacuum representation has additional physical representations.
In this particular case the extended representation (the DR-field algebra
representation) can be incorporated into the Thirring model Lagrangian, but
there is no argument that the extension always permits a Lagrangian description.

At the time of Jordan's ``neutrino theory of light'' (see the first section)
particles were simply identified with vectors obtained by applying the Fourier
components of fields to the vacuum and for bound-states one used a
covariantized quantum-mechanical description. Apart from the very peculiar
Fermion/Boson relation in d=1+1 spacetime dimension, Jordan's theory was
inconsistent with QFT. 

However if one makes use of the fact that the omnipresent vacuum fluctuations
couple all one-particle vectors with vectors obtained from the vacuum by
applying localized operators which have a nonvanishing component with the same
quantum numbers (charge, mass, spin), then Jordan's idea loses its ``crazy''
aspect (but unfortunately also some of its potential usefulness). In the
electro-weak theory one can of course find composites of the neutrino field
which fulfill these requirements and the application of scattering theory
reveals that (apart from complicated normalization factors) they are indeed
interpolating fields for photons i.e. their suitably defined large time limits
are free photon fields. Jordan's problem was an academic one, there was never
any need to have a quantum mechanical bound state description of photons in
terms of neutrinos. 

However the conceptual difficulties which come with such a bound-state picture
return with full force when we say that, e.g., a meson is made-up of two
quarks. In that case we really want to know in what sense an old-fashioned
bound state picture could make (approximate) sense in the presence of vacuum
polarization. Presently we are using the magic protection of the word ``quark
confinement'' in order to legitimize an ''effective'' return of the quantum
mechanical bound state picture which we hope to understand in a future theory
of confinement. Perhaps the meaning of ``bound state'' or ``made up'' in
quantum chromodynamics does not go beyond the message drawn from Jordan's
neutrino example, namely it refers to the fusion of (confined) charges and
should not be understood in the sense of binding of (confined) particles.

Scattering theory and the S-matrix played a crucial motivating role for
passing from pointlike fields to the setting of LQP. The Haag-Ruelle
scattering theory \cite{Haag}\cite{Araki} showed that the LSZ scattering
theory \cite{LSZ} can be derived from the principles of QFT (assuming the
presence of mass gaps) and the S-matrix stays the same if one passes from one
interpolating Heisenberg field to a composite field which acts cyclicly on the
vacuum. In particular the local equivalence class of a given field (the
Borchers class \cite{S-W}) is associated with the same S-matrix. This
insensitivity of the S-matrix against local changes was also seen within the
perturbative Lagrangian formalism, but the proofs tend to be highly technical.
The LQP formulation in terms of nets of operator algebras was the intrinsic
way to express this since relative local fields generate the same net. The
generalization of scattering theory to general operators (taken from e.g.
double cone localized algebras in the net) presented no new problems
\cite{Araki}.

Heisenberg \cite{Hei} was the first to see that the S-matrix is a global
object which poses no short ultraviolet problem since the particles
participating in the scattering process are ``on-shell'' i.e. avoid the
dangerous regions of short distance fluctuations. Hence if it is possible to
avoid fields i.e to find a calculational scheme which stays on-shell, one
could have a finite theory of interacting relativistic particles. The only
properties which Heisenberg required of the unitary S-operator besides
Poincar\'{e} invariance was a factorization property in case the wave packets
of the particles involved become spatially separated. This is a special case
of what we nowadays call the cluster property; it is automatically taken care
of if the S-matrix is that of an underlying QFT. In this way Heisenberg
formulated the first attempt for a pure S-matrix theory.

It is quite interesting to notice that Poincar\'{e} invariant S-matrices which
cluster can be constructed in a systematic way by a method which was not
available to Heisenberg. One again starts from tensor products of Wigner
one-particle spaces but this time one uses this setting in order to formulate
a quantum mechanical relativistic problem with a fixed number of particles
without vacuum polarization. The quantum-mechanical localization in the sense
of the Born probability interpretation is called the Newton-Wigner
localization; it is neither covariant nor causal\footnote{This fact caused
Wigner's disapointment and made him believe that QFT cannot satisfactorily
describe particles (private communication by Haag).}. However it acquires
these two properties asymptotically for large timelike distances and this is
the reason why it is used in time dependent scattering theory; the Haag-Ruelle
vectors at finite times depend on the Lorentz frame but the S-matrix which is
defined in terms of their large time limits is frame-independent. 

Relativistic quantum mechanics starts from a tensor product of two Wigner
particle representation and defines an interaction by going into the center of
mass frame and modifying the center of mass Hamiltonian (=invariant mass) by
adding an interaction potential which commutes with the Poincar\'{e}
generators of the full two-particle system. The mass operator in the
3-particle system has to be determined by a cluster argument; one first
transforms to the relative center of mass system of the various two-particle
subsystems where one can use the original two-body interaction. Transforming
back into the 3-particle center-of-mass frame one obtains a situation in which
the third particle is a ``spectator'' which physically corresponds to a
situation where it is spatially separated by an infinite distance from the
2-particle subsystem. Adding up the various two-particle pair interaction
operators of these spectator situations, one finally obtains the minimal
generators of the 3-particle Poincar\'{e} group in the presence of
interactions. Non-minimal 3-particle interactions are obtained by adding to
the 3-particle mass operator a term which corresponds to a direct 3-body
interaction term (which vanishes if any one of the particles becomes a
spectator). In the presence of 4 particles even the minimal implementation of
clustering yields a 3-body interaction which is uniquely determined in terms
of the original 2-body interaction. 

It is evident from this iterative construction of n-particle interactions that
the implementation of cluster properties requires the use of transformations
between different Lorentz systems i.e. in contrast to field theoretic cluster
properties, the cluster properties of a relativistic particle theory with
``direct interactions'' \cite{C-P} is frame dependent. However this dependence
drops out in the asymptotic limit i.e. for the Moller operators and the
resulting S-matrix\footnote{This framework of relativistic direct particle
interactions is being successfully used in meson-nucleon physics at energies
where relativity matters but particle creation is limited to a few channels
\cite{Poly}. Since strictly speaking it is a pure S-matrix scheme, quantities
such as formfactors of currents have to be added by hand (taking hints from
QFT) i.e. its use is strictly phenomenological.}. Therefore theories obtained
in this way are Poincar\'{e} invarinat clustering S-matrix theories par
excellence in the sense of Heisenberg's proposal.

The two-fold use of the Wigner theory as a point of departure for QFT as well
as for relativistic QM is intimately related to the existence of two different
kinds of localizations: the modular localization and the Newton-Wigner
localization \cite{N-W}. It is interesting that the two different localization
concepts have aroused passionate discussions in philosophical circles as
evidenced e.g. from a title like ``Reeh-Schlieder defeats Newton-Wigner'' in
\cite{Halvor2}. As it should be clear from our presentation, particle physics
needs both, the first for causal (non-superluminal) propagation over finite
distances and the second for scattering theory (where only asymptotic
covariance and causality is required but where the dissipation of wave packets
is important).

There is another historical reason why I have dedicated much space to recall
theories based on particles rather than fields. This has to do with Dirac's
alternative particle oriented method as compared to Jordan's quantization of
wave-fields. There simply exists no alternative to QFT if one wants to
maintain the vacuum properties. Hole theory as a theory of particles
interacting with the electromagnetic field is not consistent; the relation to
particles in the presence of vacuum polarization can only go through large
time LSZ asymptotes i.e. there can be no particles at finite times. This is
the reason why the Dirac-Fock-Podolski multi-time formulation cannot serve as
a starting point for an alternative formulation of nth order QED perturbation
theory\footnote{This explains why one does not find an n$^{th}$ order
renormalization calculation in the setting of hole theory.}, although with a
bit of artistry and hindsight one can obtain correct low order results based
on hole theory \cite{Heitler}. The scheme of Heisenberg was too general and
physically bloodless; the subsequent progress on how to extract unambiguous
results from the ill-defined perturbation theory of QED via renormalization
directed attention away from S-matrix back to QFT.

The important distinction of Heisenberg S-matrices from those resulting from
QFT is the crossing property which is related to causality and the ensuing
vacuum fluctuation properties (which allows no natural formulation in a scheme
of direct particle interactions).

Despite the great success of renormalization theory with respect to its
experimental verification, not everybody was happy. There was on the one hand
Dirac who could not reconcile the extreme technical and artistic formulation
of the renormalization recipes (since he made important contributions to
certain concepts such as the electron selfenergy we can however safely assume
that he understood what was being done) with his sense of esthetics,
mathematical rigor and elegance. We do not know what Jordan's reaction may
have been, but reading the text of his Kharkov talk one gets the impression
that he may have been a bit disappointed by the very conservative nature of
renormalized perturbation theory since he believed that only very
revolutionary conceptual changes and additions could resolve the problems
described in his talk. The loss of faith in post QED perturbative field theory
came from a growing number of physicists in the meson-nucleon physics
community who realized that the perturbative methods are inadequate for strong interactions.

This led finally to a return of S-matrix theory, but this time it was enriched
with the crossing property, an important additional requirement which was
abstracted from QFT \cite{Chew}. Looking at subsets of Feynman diagrams, it is
not difficult to see that the connected parts of on-shell formfactors or
scattering processes\footnote{The S-matrix is a special formfactor, namely the
identity operator sandwiched between outgoing bra and incoming ket states (the
opposite convention can also be used).} for different numbers of
(anti)particles in the outgoing bra and incoming ket states are related by a
process of crossing. In this process one takes away one incoming particle from
the ket state and adds it to the outgoing bra configuration with conjugate
charge (antiparticle) and opposite 4-momentum. Here a 4-momentum on the lower
real part of the complex mass shell is defined by analytic continuation from
physical values, i.e. crossing cannot be a symmetry in the sense of Wigner.
The crossing property for the connected part of formfactors of localized
operators is very suggestive in the LSZ setting; the necessary analyticity
properties follow in many cases from Lehmann-Dyson type of spectral
representations \cite{Todorov}. There is however an important fine point in
that in an on-shell setting one requires the analytic path connecting the
original with the crossed configuration to be a path which stays within the
complex mass shell. This makes crossing as it is needed in an on-shell
constructive approach (see later) a quite subtle property \cite{Bros}.

This S-matrix program was closely linked with the adaptation of Kramers-Kronig
dispersion theory to QFT. Originally it was thought of as having the same
physical content as the quantum field theoretical construction of the
S-matrix, but later it was formulated as a setting which is in opposition to
the locality concepts of QFT. The opposition finally took the extreme form of
a cleansing rage against QFT and culminated in the doctrine that the
``S-matrix bootstrap program'' has a unique solution and constitutes a theory
of everything (a TOE except gravity). 

This strange episode probably would have faded away on its own since the gap
between the wild claims and expectations of its protagonists and the physical
reality became insurmountably large, but the strong return of QFT in the form
of (nonabelian) gauge theory shortened this process. The only relic which
remained from this bootstrap program is the Veneziano dual model which later
was reformulated and extended into modern string theory (again with the
pretence of a TOE, this time including gravity). The latter does however not
represent a solution of the original S-matrix program with the field theoretic
crossing property i.e. it is not the result of an extension of existing
particle physics but rather represents an invention without known connection
to the experimentally secured physical principles underlying Jordan's
quantization of wave fields. This will become clearer in the next section
where we will show that the S-matrix with the crossing property plays a
pivotal role in the implementation of classification and construction of wedge algebras.

An important hint in favor of this new operator algebra-based approach comes
from the unicity theorem of the inverse scattering problem in LQP \cite{S3}.
It states that an S-matrix with crossing admits at most one QFT in the LQP
setting i.e. it cannot happen that a given S-matrix with crossing admits
several inequivalent QFTs. Besides the Noether currents (or rather their
algebraic counterparts \cite{Dop}) it does not distinguish particular
pointlike objects affiliated with the net of operator algebras. This complies
with the fact that in passing from classical to quantum field theory, fields
(except Noether currents) lose their individual nature since they are not
directly measurable; their main role is to provide interpolating objects for
the in/out going particles. 

Contrary to what was believed at the time of the bootstrap approach, the
S-matrix does contain local information and it is hidden behind crossing. A
conceptually more appropriate way of stating this new insight is to emphasize
that in LQP the S-matrix acquires in addition to its scattering interpretation
an important new role as a relative\footnote{The antilinear involutive Tomita
S-operator of the interacting wedge algebra differs from that of the
corresponding incoming algebra by the scattering matrix (see next section).}
modular invariant of wedge algebras which are the building blocks of LQP.

The careful shift of emphasis from Jordan's quantized wave fields to Haag's
algebraic nets (which we use in this essay in order to convince the reader
that Jordan's radical expectations have meanwhile found their appropriate
conceptual and mathematical basis) should not be misunderstood as diminishing
the importance and usefulness of pointlike fields. There is little doubt that
physically relevant models of LQP have point- or string- like covariant
generators. What is however delicate and problematic is the use of such
singular coordinatizations in (perturbative) Lagrangian calculations. It is
evident that a scheme whose only conceptual and mathematical tool consists of
pointlike fields has no alternative to the implementation of local interaction
than by locally coupling free fields i.e. by causal perturbation theory. But
it is also evident that this brings in all those technical aspects of
renormalization in which Dirac rightfully could not see neither beauty nor
rigor. 

The LQP setting offers other possibilities since e.g. a wedge algebra may be
generated by extended objects which are not identical to smeared fields
(smeared with wedge-supported test functions). The sharper localized operators
may originate from algebraic intersections and the pointlike objects may only
come into existence at the end, after one already constructed nontrivial
double cone algebras (with pointlike localization-cores). Intersections can of
course be trivial; in that case we would say that the LQP with the wedge
structure from which we started does not exist. In the causal perturbation
theory based on coupling pointlike free fields the candidates for nonsensical
theories are those which as a result of their bad ultraviolet behavior cannot
be renormalized in terms of a finite set of parameters. In that case the
perturbative delineation of models would follow the logic of the power
counting law for the interaction density\footnote{In models which contain
Fermions and Bosons one can sometimes tune the parameters in such a way that
the highest divergences compensate (supersymmetry). But the wave function
renormalization (integral over Kallen-Lehmann spectral function) can never be
finite i.e. there are no completely ultraviolet finite interacting theories in
d=1+3. For conformally invariant theories the rigorous proof is actually quite
simple \cite{S2}. The so-called finite N=4 supersymmetric gauge theories may
lead to a fake (gauge dependent) finite wave function renormalization in an
indefinite metric formulation, but the problem of infinities returns if one
constructs the gauge invariant composites.}. The achievement of
renormalization theory was that the ultraviolet divergence problems in the
presence of interactions (which started to become noticed shortly after the
discovery of quantization of wave fields and which led to despair for almost
two decades) were finally solved. To be more precise, the divergence problem
had been partially solved since it was unclear whether the power counting law
which required the operator dimension of the interaction density (which is the
sum of the dimensions of the fields participating in the interaction) to be
bounded by the spacetime dimension is an intrinsic property following from a
perturbative implementation of the principles of QFT or whether it is a
technical property of this particular implementation of perturbation theory.
This is of course an academic question as long as we cannot think of a
different approach, but as will be outlined in the next section, such an
alternative approach already exists for a special nontrivial family of d=1+1 QFTs.

The following illustration may serve as a warning against interpreting the
renormalizability criterion in terms of power counting too naively as a
well-defined procedure to screen between existing and non-existing QFTs.
Suppose we are interested in interacting massive vector mesons. The easiest
way to describe free massive vector mesons is by a transverse vector potential
$A_{\mu}$ which obeys the Klein-Gordon equation. But since contrary to
classical expectations its operator dimension is $d_{A}=2$ (any other
covariantization of the Wigner representation would only increase the operator
dimension), and since each interaction is at least trilinear in free fields,
the interaction density of any model involving physical vector potentials has
inevitably operator dimension $\geq5;$ which would make it nonrenormalizable
i.e. the dimension of the interacting field would keep increasing with the
perturbative order and these higher polynomials would cause an ever-increasing
number of independent parameters. 

At this point one would be reminded that massive vector mesons should be
treated as gauge theories with the mass being generated by a Higgs mechanism.
But let us ignore this advice and use instead a cohomological extension of the
Wigner one-particle theory of a massive vector meson. By this we mean an
enlarged unphysical (i.e. with indefinite metric ghosts) representation of the
Poincar\'{e} group with a nilpotent $\delta$-operation such that the original
Wigner space is the cohomology of the extended situation. This is possible and
the associated Fock space formalism is a particularily simple linear
realization (bilinear in the Lagrangian) of the BRST formalism which is only
possible in the massive case \cite{D-S}. The unphysical field has formally
operator dimension d=1, and therefore its coupling becomes renormalizable in
the new power counting. The cohomological nature of the extension gives us the
hope that as a benefit of deformation stability of cohomology we will be able
to descend to a physical sub-theory. This is indeed the case but only after
realizing that perturbative consistency of the renormalized theory requires
the introduction of new physical degrees of freedom whose simplest realization
is a scalar field. This is of course the Higgs field except that in the
present case it has a vanishing vacuum expectation; the theory was massive to
begin with, and there is no mass-generating role which would require a Higgs
condensate. The physical expectation values are identical to those obtained
from the gauge theoretical construction together with the Higgs mechanism. 

The so obtained physical theory (after the cohomological descent) has no
memory about the cohomological trick; the procedure is reminiscent of the use
of a catalyst in chemistry in that the cohomological extension was used in an
intermediate step and again removed through the descent \cite{D-S}. This trick
has stabilized the operator dimension d=2 of the physical vector meson modulo
logarithmic corrections and (in contrast to the naive approach) did not
require to introduce an infinite set of new parameters 

The interest of this illustration is not the result itself, but rather the
different way in which it is obtained (some of the following ideas can be
traced back to extensive work of Scharf and collaborators \cite{Scharf}). It
shows, as was always expected by some physicists, that the approach via gauge
theory and Higgs mechanism of mass generation has no intrinsic physical
meaning (but rather represents a useful technical tool for doing
computations). In contrast to a classical field theories with vector fields
for which the gauge principle is needed to characterize Maxwellian
interactions, the quantum field theoretical renormalizability of massive
vector mesons fixes the interaction all on its own uniquely in terms of one
coupling strength. The semiclassical reading of that unique vector meson
theory reveals of course that it looks exactly like the quantization of
classical gauge theory with the Higgs mechanism. Since quantum principles
(even if they are presently insufficiently understood) are more basic than
classical principles, one should perhaps turn around the historical approach
in favor of the opposite reading%

\[
renormalizability\overset{semiclassically}{\longrightarrow}%
classical\,\,gauge\text{ }principle
\]
since a theory which is already unique in the perturbative setting of
renormalizability does not need any further selection principle. Pragmatically
speaking we have only reduced the somewhat mysterious classical gauge
principle (for which we have beautiful differential geometric presentations,
but geometry is no substitute for causal quantum physics) to the mystery of a
renormalizability principle (of which we still lack an understanding within
the beautiful setting of LQP). But whereas we still have a future chance to
unravel the latter, one does not expect that anything can be added to the
former. The strategy indicated works strictly speaking for massive vector
mesons, one expects that in the massless limit the scalar particles will
decuple (possibly coupled with the appearance of new charges).

This illustration also suggests that perhaps the nonrenormalizability
statement concerning the coupling higher spin s%
$>$%
1 fields should be taken with a grain of salt. To be sure, if we start with a
physical free field with higher spin, the short distance scale dimension is
certainly beyond the canonical value one, which means that the perturbative
iteration will increase the short distance divergences by inverse powers which
increase together with the number of new parameters with the perturbative
order. But there may exist a cohomology-like magic which leads to an
unphysical but renormalizable situation such that after a descent to a
physical subsystem the operator dimension of the higher spin Heisenberg field
is stabilized around its initial physical free field value modulo logarithmic
corrections. According to the above considerations this is precisely what
happens for massive vector mesons, but since the systematics of such tricks is
not known, the status of higher spin representations in interactions is
unclear. This brings us back to Jordan's 1929 critical assessment of the
situation and the persisting need for a radically different approach as
compared to the standard quantization formalism, even after the discovery of renormalization.

There have been attempts to improve the short distance behavior in the
conventional point-like field setting by enforcing a partial cancellation (see
previous footnote) of divergencies between Fermions and Bosons via the
so-called supersymmetry. But the tuning between Fermions and Bosons which
achieves this supersymmetry has some consequences which show that this
supersymmetry, unlike any other symmetry, is a very peculiar kind of symmetry.
Whereas the coupling of systems with normal symmetries (internal, Lorentz) to
a heat bath causes a sponteneous breaking, the supersymmetry ``collapses''
\cite{B-O}. This means that there is no enlargement of the Hilbert space by
combining different values of the breaking parameter (e.g. the preferred
rotational direction in the case of a ferromagnet) such that the symmetry can
be recovered (at the expense of the cluster property) in the larger reducible
representation space as is the case for spontaneous breaking. Another more
serious reason is that the modular method applied to observable algebras
(which are neutral, i.e. on which inner symmetries act trivially and spacetime
symmetry is described by the Poincar\'{e} group and not its covering) reveals
all inner symmetries but, there is no indication for supersymmetry. Since it
goes against the observable algebra--charged field algebra dichotomy by mixing
observable fields with fields carrying the unimodular charge, a follower of
the LQP logic may say what God has separated men should not force together. In
fact in the framework of perturbative study of renormalization group flows in
a multicoupling setting, supersymmetry appears as an accidental point at which
certain short distance compensations take place. This special situation looks
somewhat accidental (which corroborates with the previous failure of
supersymmetry to show up in the modular setting). But of course these
arguments do not convince somebody who wants to see a supersymmetric partner
behind each particle (at least for asymptotically large energies). 

The LQP approach certainly satisfies Jordan's requirements which he formulated
in his Kharkov talk and as a result of its mathematical precision and
conceptual beauty it may even have pleased Dirac. But the crucial question is
whether it is capable of leading to a classification and construction of
models which can explain important results in particle physics. There is
reason for optimism, but the research is very much in its infancy. In the last
section I will try to present some constructive ideas and a few encouraging results.

\section{LQP and ultraviolet-finiteness}

As soon as one decides to work with pointlike fields, it is very hard to think
of a construction scheme which is different from the standard way of coupling
free fields and using the Wick-ordered interaction density for the iterative
computation of the renormalized time-ordered products via their vacuum
expectation values. The only notable exception is chiral conformal theory
which deals with fields which are localized on a lightray. Although such
models are technically speaking not identical with free fields, they are (like
free fields) uniquely determined in terms of their commutation structure
(which in turn is fixed by braid group representations and possibly additional
combinatorial data); with other words there are no deformable interaction
parameters (coupling strengths). Although the complete descriptions in terms
of n-point correlations (or explicit formulas in terms of auxiliary free
fields) have only been constructed very special cases, the known facts makes
them very interesting objects of field theoretic research. The methods of
constructions of pointlike fields are representation-theoretic
(representations of ``Lie-field'' algebras of the current-algebra or W-algebra
type) and leave no room for ultraviolet divergencies since the scale dimension
is part of the spacetime group theory. The operator algebraic approach and the
pointlike formalism turn out to be equivalent \cite{Joerss}. Chiral theories
arise as ``conformal blocks'' in a decomposition theory for 2-dimensional
local (bosonic or fermionic) conformal field theories. This origin of this
decomposition is basically group theoretical. 

The relevant space-time symmetry group is the universal covering of the
conformal group and the relevant ``living space'' is not Minkowski spacetime
but rather the universal covering of compactified Minkowski spacetime which
carries its own global causality structure \cite{Mack}. 

A more convenient description in terms of operator distribution valued
sections on the (compactified) Minkowski spacetime (as the base of a vector
bundle) can be obtained by decomposing the global covering field according to
the center of the conformal covering \cite{S-S}. The resulting central
components (conformal block fields) are the desired sections. For the special
case that the central decomposition is trivial (no complex phase
factors\footnote{The decomposition theory shows that the phases of the central
phase factors are linear combinations of the (anomalous) scale dimensions
which occur in the theory \cite{S-S}.} under central transformations) the
field is a free field which lives on the compactified Minkowski space (or on
its two-fold covering in case of an odd number of spacetime dimensions) and
therefore fulfills the quantum version of Huygens principle i.e. the
(anti)commutator has its support on the lightcone. This happens e.g. for free
photons. Jordan was aware of the validity of this principle in his quantized
wave field setting \cite{cau}, but not of the fact that a conformal
interactions gives rise to an algebraic modification in the Huygens region
\cite{S1}. 

Conformal invariant theories offer presently the best chances for explicit
classifications and constructions since the mechanism for interactions is
fully accounted for by the algebraic structure in the Huygens (timelike)
region (which in turn determines the spectrum of scale dimensions) i.e. the
Jordan-Haag credo of an intrinsic access to local quantum physics without
classical crutches is fully realized. There is no classical counterpart of the
timelike ``reverberation'' structure for even spacetime dimension; the
covering group representation aspect is an extremely rich generalization of
the spin phenomenon i.e. one has the analogy%
\[
spin\text{ }(spacelike)\backsim anomalous\,\,dimension\,(timelike)
\]
The best studied case is that of two-dimensional conformal theories where the
conformal group tensor factorizes into two Moebius groups with the center of
the covering following suit. This leads to the very peculiar tensor
decomposition of the two dimensional conformal QFT into two chiral theories.
The restrictiveness of the resulting algebraic structure together with the
acquired mathematical knowledge about Kac-Moody- and diffeomorphism- algebras
permitted Belavin Polyakov and Zamolodchikov \cite{BPZ} to discover the first
nontrivial family of chiral models which they called ``minimal'' models since
their representation structure was completely fixed in terms of the
commutation relations of only one field, which was the energy-momentum tensor.
Whereas previous model illustrations of the block decomposition theory
consisted of exponentials of free massless Boson fields (which appear
naturally in the solution of the massless Thirring model) and one could not
have been sure that the apparent generality of the decomposition can be
realized by models, the discovery of the minimal models showed the enormous
scope of its content and the richness of the ensuing anomalous dimension spectrum.

The above mentioned algebraic structure in the Huygens region for d=1+1
conformal theories splits into a separate algebraic structure of the left and
right moving chiral components, i.e. it becomes re-processed into the two
lightray commutation structure. Since lightlike can also be seen as a limit of
spacelike, the chiral commutation structure can be interpreted in terms of
braid group statistics (of fields). This viewpoint turned out to be very
useful for classification and constructions of chiral theories \cite{R-S}%
\cite{Fro}. In fact the ultimate goal in chiral QFT is to obtain a complete
classification and construction of the interval-indexed chiral nets or their
spacetime pointlike field generators based on braid group statistics (and
possibly additional combinatorial data). The presence of the powerful
Tomita-Takesaki modular theory of operator algebras which has been
successfully adapted to LQP \cite{Borchers} nourishes hopes, that this goal
may be reached in the not too distant future.

The chiral theories are much more than a training ground for nonperturbative
methods in QFT. Since physically speaking they arise as zero mass limits of
massive theories, one expects that they still retain physical informations
about highly inclusive scattering processes. But in view of the fact that
standard scattering theory is not applicable (the LSZ limits in interacting
conformal theories vanish \cite{S2}) and a formulation which aims directly at
probabilities for inclusive processes has not yet been worked out to the
extend that it can be directly applied \cite{Haag}, this still remains a
problem of future research. It seems to be undeserved luck that precisely
chiral theories, about which one has the presently most detailed knowledge,
can be used for higher dimensional LQP. This is because it can be shown that
chiral theories are the building blocks in the holographic lightfront
projection of any local QFT (independent of the spacetime dimension)
\cite{LHF}. 

The idea leading to this mathematically and conceptually very precise
formulation of holography in the setting of LQP is the following. One starts
with the Unruh situation of a (Rindler-) wedge restricted QFT which
automatically inherits the Hawking thermal aspects\footnote{The vacuum state
resricted to the wedge algebra becomes a thermal KMS state at the Hawking
temperature.
\par
{}} \ The linear extension of the (upper) causal horizon is a lightfront which
inherits a 7-parametric subgroup of the 10-parametric Poincar\'{e} group. The
lightfront with its inherited unusual causal structure is very different from
globally hyperbolic spacetimes, but precisely this makes it a powerful
instrument for the exploration of QFT associated to the ambient Minkowski
spacetime. With the exception of d=1+1 (where one needs data on both
lightrays) the classical data on the lightfront determine the global ambient
theory. Even more, the data on the wedge (upper) horizon determine the data in
the associated wedge. The smallest regions on the lightfront which still cast
ambient causal shadows are semiinfinite strips with a finite transverse
extension on the lightfront. Compact regions on the lightfront do not cast any
ambient shadows; in fact they do not even cast shadows within the lightfront
(the manifestation of its extreme deviation from hyperbolic propagation). 

The marvelous aspect of LQP consists in enabling us to transfer these
structures to the local quantum realm; the algebra localized on the horizon is
identical to the global wedge algebra, but the net substructures of both
algebras are very different (apart from the equality of the before mentioned
semiinfinite strip algebras with their ambient causal shadows). For
interaction-free theories these claims can be verified by standard methods
i.e. by restricting the pointlike fields to the lightfront. For interacting
fields, which in d=1+3 necessarily lead to Kallen-Lehmann spectral functions
with affiliated infinite wave function renormalization factors, the lightfront
theory cannot be obtained by restriction. In this case one has to use pure
operator-algebraic modular methods. It turns out that the lightfront net has
no transverse vacuum fluctuations, all vacuum polarization is compressed into
the longitudinal lightray direction. This complies perfectly with the
7-parametric symmetry group aspect; the holographic projection of the two
``translations'' contained in the 3-parametric Wigner little group of the
lightray in the lightfront precisely accounts for the transverse Galilei
invariance of a fluctuationless transverse quantum mechanics. 

The lightfront QFT is the only known case in which QM for a subsystem returns
in the midst of QFT without any nonrelativistic approximation. In lightray
direction one finds a Moebius invariant chiral theory. The rotational subgroup
of the Moebius group does not result from the holographic projection of the
Poincar\'{e} group (i.e. is not a subgroup of the mentioned 7-parametric
group), rather it results from ``symmetry enhancement'' \cite{Verch} on the
lightfront\footnote{Since in chiral theories the diffeomorphisms of the circle
turn out to be of modular origin \cite{F-S2}, one expects the higher
diffeomorphisms to be also symmetries of the holographic projection. It turns
out that these transformations act in a fuzzy way on the ambient theory.
\par
{}}; the phenomenon of symmetry enhancement and its connection to universality
classes is well known from the critical limit (massless limit, scaling limit).
The difference with the lightfront holography and its associated universality
aspect is that the ambient theory and its holographic projection live in the
same Hilbert space; in fact the two theories are coming from identical global
algebras, their difference is only in the spacetime net indexing of
subalgebras which causes a relative nonlocality. Therefore the new symmetries
which are represented by diffeomorphisms of the holographic projection have
already been present in the ambient theory. They were not perceived as
symmetries simply because their action on the ambient theory is ``fuzzy'' i.e.
not presentable in terms of geometric transformations (but only as
support-maintaining transformations of test function spaces if one uses the
setting of pointlike fields). Whereas the holographic process from the ambient
theory to its lightfront image is unique, the holographic inversion is not; it
remains an interesting problem to classify the class of ambient theories
having the same image or adding additional informations to the holographic
projection which makes the inversion unique. Presently lightfront holography
is not in a state where it could be useful for the construction of ambient QFTs.

The holographic approach becomes more favorable if one restricts ones interest
to problems which do not require to solve the holographic inversion, but
permit to be studied in the holographic projection. The most prominent such
problem is the assignment of an entropy to the phenomenon of thermalization
caused by causal localization. A special case is Hawking's black hole
thermalization which through the imposition of the classic thermodynamic basic
laws of heat bath thermality leads to Bekenstein's area law for the assigned
entropy. The fundamental problem behind this pivotal observation is to find a
direct derivation of a quantum entropy in the setting of localization-caused
thermal aspects which applies not only to black holes but also to black hole
analogs\footnote{The termal aspects of localization do not depend on the
presence of spacetime curvature but rather on an extension of the idea of
``surface gravity'' which sets different scales for true gravity theories from
those of their acoustical, hydrodynamic or optical analogs.}. 

An excellent testing ground is the restriction of the global vacuum to the
wedge algebra or that of its causal horizon i.e. the Unruh-Rindler thermal
situation. The relevant KMS operator is the Lorentz boost which is not of
trace class and hence permits no definition of entropy. This is a general
property of all localized algebras; the restricted vacuum leads to (generally
non-geometric) KMS states which cannot be tracial since sharply localized
algebras are of hyperfinite von Neumann type III$_{1}$ which simply do not
have tracial states, hence a notion of entropy cannot be defined. The physical
origin of this phenomenon are the uncontrollably big vacuum fluctuations which
accompany the creation of sharply defined causal localization boundaries. The
remedy is to make a fuzzy surface which leaves the vacuum polarization to
reorganize themselves in a ``halo'' surrounding the original localization
region. This is done with the help of the so-called split inclusion property,
which, different from a cutoff, maintains the original local theory while only
reorganizing some local degrees of freedom in the finitely extended halo
(whose extension is a control parameter for the size of vacuum fluctuations). 

The vacuum state restricted to the split inclusion becomes a thermal density
matrix (with the Hawking temperature) in an appropriately defined tensor
factorization of the total Hilbert space. This sequence of density matrices
for decreasing halo size converges (as expected) against the dilation operator
(the holographic image of the wedge-associated boost) which is a non-trace
class operator which sets the thermal KMS properties of the Unruh-Rindler
effect; thus if one would be able to associate a split-localization entropy
with the finite halo situation, one can be sure that this is naturally
associated with the Hawking-Unruh temperature aspect. The transverse symmetry
of the horizon (whose linear extension is the lightfront) forces the concept
of an area (the dimension of the transverse edge of the wedge) density of
split-localized entropy. 

This shows that the prerequisites for a Bekenstein-like quantum area law for
localization entropy are met in a surprisingly generic manner. But for the
control of a limiting halo-independent area density two more properties remain
to be established:

(1) The increase of the area density with decreasing halo size $\varepsilon$
is universal (model calculations indicate that it goes like $ln\varepsilon)$
so that it possible to have a finite relative area density between systems
with different quantum matter content.

(2) The validity of thermodynamic basic laws for causal localization-caused
thermal behavior which parallels those of the standard heat-bath thermal setting.

These problems are presently being investigated.

The technical advantage of the holographic approach for particle physics lies
in the simplicity of chiral QFT, but the unsolved problems of the holographic
inversion prevent presently the return to the particle setting of the ambient
original theory (interacting conformal theories do not permit a Wigner
particle interpretation \cite{S2} and an associated scattering theory; for
this reason they are not of \textit{direct} physical relevance to particle physics).

Fortunately there are constructive ideas which stay within the S-matrix
particle framework. Their basis is the observation that the scattering matrix
in LQP admits an interpretation in terms of the modular data of wedge
algebras, which is characteristic for causal localization (and the ensuing
vacuum polarization and TCP symmetry) property. This can be seen by combining
two known facts. On the one hand the $\Theta\equiv TCP$ invariance of the
$S-$matrix \cite{S-W} can be written in the form%
\[
\Theta=\Theta^{in}S
\]
where $\Theta^{in}$ is the TCP operator of the free field theory for the
incoming fields. The second fact is that the Tomita $S_{T}-$operator for the
wedge algebra $\mathcal{A}$($W_{0}$) relative to the vacuum state vector
$\Omega$ which is the antilinear closable operator defined by
\[
S_{T}A\Omega=A^{\ast}\Omega,\,\,A\in\mathcal{A}(W)
\]
has a polar decomposition (choosing the reference wedge $W_{0}$ as $x>\left|
t\right|  $)%
\begin{align*}
&  S_{T}=J\Delta^{\frac{1}{2}}\\
J &  =\Theta U(Rot_{x}(\pi)),\,\,\Delta^{i\tau}=U(\Lambda_{x-t}(2\pi\tau))
\end{align*}
where the ``radial'' part $\Delta^{\frac{1}{2}}$of the polar decomposition is
the positive operator $e^{-\pi K}$ with $K=$generator of x-t boost and $J$ the
antilinear involutive ``angular'' which is apart from a $\pi$-rotation around
the x-axis identical to $\Theta.$ Since the S-matrix commutes with the
connected part of the Poincar\'{e} group, the relation between $J$ and $S$ is
the same as that with $\Theta$ and $S$ namely\footnote{As in most derivations
involving the S-matrix we assume asymptotic completeness $H=H^{in}.$}%
\[
J=J^{in}S,\,\,J^{2}=1
\]
This is a special case of the Tomita Takesaki modular theory for operator
algebras whose aim is to classify and characterize pairs of operator algebras
and cyclic and separating state vectors ($\mathcal{A},\Omega$) in terms of
their modular operators namely an antiunitary involution $J$ (generalized
``TCP'') and a one-parametric unitary modular group $\Delta^{it}$ (a
generalized boost ``Hamiltonian'' $\Delta^{it}=e^{2\pi itK}).$ The Tomita
operator $S_{T}$ has characteristic properties which make it a unique object
of mathematical physics\footnote{It does not appear outside modular theory and
therefore cannot be found in past and present mathematical physics
textbooks.}, it is an anti-unitary operator which is involutive on its domain
$S_{T}^{2}\subset1$ and ``transparent'' i.e. its range equals its domain. It
is the only object which is capable to encode geometric properties into domain
properties; modular theory is the key for understanding the connections
between the operator aspects of LQP and their geometric manifestations (see
more remarks below). In an essay about Jordan and his legacy we cannot do more
than make an evocation of this rich mathematical theory in terms of some of
its physics adapted formulas; the non-expert reader may consult
\cite{Borchers} in order to learn about its deep content.

Modular theory of operator algebras is helpful and often indispensable for an
intrinsic analysis of QFT which avoids the (always) singular pointlike field
coordinatization forced upon us by field quantization while preserving its
physical content. The first step of such an analysis consists in extracting
informations about inner- and spacetime- symmetries from the algebraic net
structure. Here modular theory provides a unifying viewpoint; not in the sense
of a group theoretical marriage (which, as shown by O'Rafarteigh, is
impossible under reasonable assumptions) but rather by providing a common
origin: algebraic inclusions. The spacetime automorphism groups of Minkowski
spacetime (Poincar\'{e} group) and Dirac-Weyl compactified Minkowski spacetime
(the conformal group) are generated from the modular group of wedge algebras
relative to the vacuum state (in fact one only needs a finite number of
modular generators). There are indications that if one generalizes the modular
theory of chiral nets (indexed by intervals on the lightray) to
multi-intervals and permits also non-vacuum states one can modular generate
the full diffeomorphism group of the circle \cite{F-S2}.

Since the discovery of (symmetry) group theory by Galois, one has the option
of analysing inclusions\footnote{Galois studied the inclusion of a number
field in its extension obtained by adjoining the roots of a polynomial
equation with coefficients from the number field.} in order to study
symmetries. There are three types of inclusions of operator algebras: (1) V.
Jones- or DHR-inclusions, (2) modular inclusions and (3) split inclusions.

The DHR inclusions originate from the DHR endomorphisms i.e. from globally
inequivalent representations which restricted to the individual algebras
forming the net yields unitary equivalent subalgebras which have the same
spacetime indexing. On the mathematical side these inclusions belong to class
of Jones inclusions which are characterized by the existence of conditional
expectations from the larger to the smaller algebras. Modular theory provides
a geometric characterization; a necessary and sufficient condition for Jones
inclusions is that the modular group of the smaller algebra is the restriction
of that of the bigger one (the Takesaki theorem). Physicists usually met
conditional expectations in the commutative Euclidean Kadanoff-Wilson scheme
of renormalization group decimation where modular groups are trivial and
conditional expectations from the original- to the thinned out- algebra of
infinite dimensional function spaces always exist. Jones inclusions can be
encoded into group symmetries and their ``para-group'' generalizations; in the
case of 4-dimensional observable nets indexed by subsets of d=3+1-dimensional
Minkowski spacetime one obtains groups which always can be viewed as subgroups
of SU(n) for large enough n; the DHR theory confirms precisely Jordan's
results which led him to interpret inner symmetries as ``dual'' to permutation
group statistics.

The modular inclusions result from going beyond Jones inclusions, but doing
this in a controllable ``minimalistic'' way by retaining a halfsided action of
the restricted modular group on the smaller algebra. The resulting group is
isomorphic to the translation-dilation group on a line. By adding further
algebraic properties (modular intersections) obtained from the concrete
physical situation, one can extend the noncompact groups to the Moebius group
etc. These methods are used in the aforementioned algebraic lightfront
holography. The perhaps most surprising results obtained by modular methods,
which highlight the inexorable connection between relative position of
subalgebras and geometry, are those in \cite{K-W} where it is shown that a
finite set of subalgebras (3 in d=1+2, 6 in d=1+3) carefully tuned in a
precise relative position with respect to each other (as operator algebras in
a common Hilbert space) can define a full QFT; i.e. these data do not only
determine the spacetime symmetry group but also the full net structure (by
transforming with the original algebras with the symmetry group and forming
algebraic intersecting)\footnote{The algebras used in this construction are
identical (the unique hyperfinite type III$_{1}$ von Neumann factor), i.e.
like in case of a finite collection of points the information is residing
solely in the relations.}. It seems that any serious-minded approach about
connections between quantum physics and geometry (e.g. quantum gravity) should
take these observations into account. According to the considerations about
the thermal manifestations of localization at the beginning of this section,
thermal aspects should be an important part of the quantum-geometry connection.

The third type of inclusion namely the split inclusion was already alluded to
in connection with recovering quantum mechanical (inside/outside quantization
box tensor factorization) properties in LQP. Although the individual local
algebras in the algebraic net are well-defined mathematical objects, their
mathematical type III nature (which is inexorably related to the infinitely
large vacuum fluctuations at the boundary of the sharp localization region)
does not permit a tensor factorization into the algebra and its commutant. The
split property allows to recover a tensor factorization by permitting the
vacuum fluctuations to spread into a fuzzy halo of arbitrarily thin diameter;
unlike a momentum space cutoff which converts the local theory into an unknown
object, the splitting is a local process which is well defined in the original
theory and does not require to throw away degrees of freedom. It includes a
local algebra into a larger local algebra with localization region equal to
the original region plus halo and secures the existence of type I quantum
mechanical algebras with a fuzzy intermediate localization. In contrast to the
previous two types of inclusions it is not unique which corresponds physically
to the many possible interpolating shapes for the vacuum polarization clouds
in the fuzzy halo region. Both the modular groups of the two algebras with
respect to the vacuum are generally not spacetime diffeomorphisms, but rather
fuzzy transformations which in the setting of pointlike fields act on the test
functions in such a way that their support in the respective regions is kept
fix. \ 

As in the case of the Wigner one-particle spaces in the previous section, we
may define the Tomita operator in terms of the associated Lorentz boost, the
TCP operator of the asymptotic particles and the S-matrix and use it as before
to define a real subspace as the +1 eigenspace; its complexified span is the
dense subspace of modular wedge localized vectors. But different from the
one-particle case there exists no functor which encodes the spatial modular
theory into its algebraic counterpart. In looking for a substitute, one should
remember that the crossing property was essential for the uniqueness proof of
the inverse scattering problem. Hence we should add this property to the
causality and spectral requirements which define a theory within the LQP
setting and worry about its conceptual position within LQP at a later state.
There are two cases in which this would not be necessary: for formfactors in
interaction-free theories (algebras generated by smeared free fields) and the
interacting d=1+1 factorizing models. So the first step should consist in a
clarification what do we mean by interactions if we do not have the Lagrangian
``crutches'' at our disposal. This can be done with the help of
``polarization-free generators'' (PFG \cite{schroer}) i.e. operators $F$
affiliated with localized operator algebras $\mathcal{A(O)}$ which applied to
the vacuum state vector generate a one-particle state vector with no admixture
of vacuum polarization (particle/antiparticle pairs)%
\[
F\Omega=one-particle\,\ vector
\]
Such PFGs exist in case of algebras generated by free fields for any region
$O$; according to modular theory they are also available for arbitrary QFTs if
one chooses for $O$ a wedge region. Hence one defines interacting theories as
those for which no sub-wedge PFGs exist, a definition which obviously does not
use Lagrangian crutches. Let us now characterize factorizing d=1+1 models as
models for which the S-matrix is purely elastic and where the n-particle
elastic S-matrix is moreover determined by the elastic two-particle S-matrix.
The cluster properties then fix the form of the n-particle scattering as%
\begin{align*}
&  S^{(n)}(\theta_{1,.....}\theta_{n})=\prod_{i<j}S^{(2)}(\theta_{i}%
,\theta_{j})\\
&  p=m(ch\theta,sh\theta),\,S^{(2)}(\theta_{i},\theta_{j})=S(\theta_{i}%
-\theta_{j})
\end{align*}
The momentum space rapidity parametrization is very convenient to formulate
crossing and the presence of bound states\footnote{The boundstate picture for
an elastic S-matrix is a quantum mechanical one, however the S-matrix particle
hierarchy passes into the charge hierarchy of the previous section as soon as
one studies the operator-particle relation for operators localized in
sub-wedge regions.} for the meromorphic two particle S-matrices; in fact
meromorphy, unitary and crossing symmetry leads to a classification of
scattering matrices $S^{(2)}(\theta),$ with other words the bootstrap program
in this limited factorization context is completely solvable and there is no
unique solution (no theory of everything) but there are infinitely many
S-matrices with the crossing property. Following Zamolodchikov one can
associate a formal algebra with such S-matrices which fulfills the Z-F algebra
which in the simplest case is of the form%

\begin{align*}
Z^{\ast}(\theta)Z^{\ast}(\theta^{\prime})  &  =S(\theta-\theta^{\prime
})Z^{\ast}(\theta^{\prime})Z^{\ast}(\theta)\\
Z(\theta)Z^{\ast}(\theta^{\prime})  &  =S(\theta^{\prime}-\theta)Z^{\ast
}(\theta^{\prime})Z(\theta)+\delta(\theta-\theta^{\prime})
\end{align*}

A closer examination reveals that such operators are the positive and negative
frequency parts of PFGs for the wedge algebras \cite{schroer}. For the proof
of the modular properties, especially the KMS property of the generating PFG
operators, the crossing property of $S(\theta)$ turned out to be crucial, a
fact which was to be expected on the basis of the similarity between both of
these cyclicity relations. A detailed account with all mathematical rigor for
the case without bound states was recently presented in \cite{Lechner}. In the
presence of bound states the formulation in terms of algebraic commutation
relations between PFG generators is more involved since their presence changes
the structure of the Fock space which would lead to the presence of projection
operators and additional terms in the Z-F commutation relation. In that case a
definition of the action of the Z-F operators on the multiparticle Fock space
vectors is more convenient. The KMS property amounts to a compensation of the
explicit bound state contributions with terms coming from contour shifts from
S-matrix poles with new explicit antiparticle contributions coming from the
contour shifts across the crossed poles. Further restrictions beyond the
crossing symmetry in the presence of bound state pole situation do not seem to
arise from the KMS property, but a treatment on the same level of rigor as for
the case without bound states is still missing.

Knowing the wedge algebra, the remaining steps of the algebraic approach
consist in computing better localized algebras (smaller localization regions)
by forming algebraic intersection of wedge algebras in general position (by
Poincar\'{e} transforming the wedge algebra in the reference position
$W_{0}).$ This is of course a step demanding a new type of mathematical skill,
but the important point here is that nowhere in this constructive road map one
has to face a short distance problem. The process of construction stops at the
double cone algebras of arbitrarily small size whose localization-core is a
point. Whenever the resulting net of operator algebras has pointlike
generating fields (there are no known counter arguments) one may of course
coordinatize the net in terms of such fields, but in doing this one looses the
unicity and intrinsic nature of the description while possibly gaining
simplicity and familiarity of the formalism. Intersections may of course turn
out to be trivial i.e. just consist of multiples of the identity operator. In
this case we would say that there is no local theory with the required data
which would be the analog of saying that a theory in the standard setting is
nonrenormalizable or irreparably ultraviolet divergent. The method of
constructing the net of operator algebras from a reference wedge algebra is
completely general. The main difference to the standard approach is that in
the LQP setting one is not forced to work with objects which already have the
best possible localization from the beginning throughout the calculation but
rather one thinks of generating operators which are not better localized than
the algebras they are supposed to generate i.e. their localization can never
coalesce to a point.

In the case of d=1+1 factorizable models one can actually compute the
intersections thanks to the simplicity of the PFG generators. In the case of
absence of bound states one obtains a LSZ like\footnote{I am referring to the
representations of Heisenberg fields as infinite series in the incoming field
with retarded coefficient functions whose Fourier transforms have known
analyticity properties.} infinite series in Wick ordered PFG operators with
meromorphic coefficient functions with Paley-Wiener-Schwarz fall-off
properties which reflect the size of the localization region. These
coefficient functions are the connected parts of the formfactors of the double
cone localized operators i.e. the localized operators are only known as
sesquilinear forms on multiparticle states, the existence of operators behind
those infinite series remains an open problem as it did in the analogous LSZ
case (where pointlike Heisenberg fields were represented as infinite series in
terms of incoming fields). The structure of the formfactor equations from the
intersecting property is identical to those relating the different formfactors
in the approach based on Smirnov's recipe. A remarkable property of the
formfactors and the S-matrix is that they are analytic in certain coupling
strength parameters g which appear in some of the solutions around
g=0\footnote{In certain models for which a Lagrangian description is known,
these parameters coalesce with the usual coupling strengths.}. There are
however arguments that the power series for off-shell quantities as vacuum
correlation functions in the coupling strength can never be convergent (at
best summable in some weaker sense). This raises the question whether the
difference in perturbative convergence properties between on-shell and
off-shell objects could be of a general nature.

Since the contemporary literature creates sometimes the wrong impression that
in some sense the physical scope of QFT is exhausted by the Lagrangian
quantization method (functional integral representations in terms of Euclidean
actions), it may be helpful to point out that there are interesting models
which do not support such a view. This is not only demonstrated by the quantum
Wigner strings but also by many examples of d=1+1 factorizing theories (e.g.
the Koberle-Swieca Z(n) models \cite{B-K}) which cannot be ``baptized'' by a
Lagrangian name.

Unfortunately the starting point of this approach, namely the wedge generating
property of the PFG operators, is, apart from one exception\footnote{The
exceptional case consists of ``free anyons'' in d=1+2. In the Wigner
representation theory they correspond to $s\neq$ halfinteger (see previous
section). The resulting braid group statistics is incompatible with the
existence of subwedge PFGs but admits the existence of temperate PFGs for
wedges \cite{Mund2}. In other words the local generation of anyon particles as
opposed to Bosons and Fermions requires the presence of vacuum polarization
clouds, even if there is no genuine interaction (vanishing scattering cross
section).}, limited to the d=1+1 factorizing models (one needs ``temperate''
PFGs which are only consistent with purely elastic scattering \cite{BBS}).
According to \cite{BBS}, the existence of temperate PFGs in $d>1+1$ result in
the triviality of scattering (hence the adjective ``free''). It is clear that
in order to find a substitute for the non-existing temperate PFG in the
general case one has to obtain a better understanding of the mysterious
crossing property and its possible relation to modular aspects. For a step in
this direction I refer to a forthcoming paper \cite{on-shell} in which the
crossing symmetry for the connected formfactors of a special object (a
``Master'' field) admits an operator KMS modular interpretation in terms of an
auxiliary nonlocal field theory on the rapidity space.

The historically minded reader will have noticed that the present ideas about
an on-shell constructive program are closely related to the S-matrix bootstrap
approach of the 60s, a fact which was already alluded to in the previous
section. The main difference is that in the new setting the S-matrix is viewed
as a special formfactor (of the identity operator between in and out states)
and the program is extended to formfactors of localized operators with vacuum
polarization which are, via the crossing property, able to link virtual
particle creation with real creation. The aim of this new ``on-shell
approach'' is to implement the classification and construction of nets of
operator algebras within the LQP setting without falling into the trap of
ultraviolet divergencies which is an inexorable aspect of quantization schemes
and their emphasis on singular field coordinatizations. The success of the
d=1+1 bootstrap-formfactor program for factorizing models, and the natural
explanation of its recipes in terms of the modular aspects of operator
algebras shows that the new ultraviolet-finite setting is not empty. 

The faith in this program for the general case (beyond the factorizing models)
comes primarily from the uniqueness of the inverse scattering problem in
regard to the field coordinatization-independent LQP net formulation
\cite{S3}. In a situation of uniqueness one expects (for those cases were a
theory in the LQP sense really exists) to find at least in a perturbative
method for classification and construction. Such a new on-shell perturbation
theory should not only reproduce the perturbative expressions for standard
renormalizable theories in the off-shell Lagrangian formulation upon
restriction to on-shell formfactors, but also permit to treat interacting
massive vector mesons without ghosts and with the perturbative presence of the
scalar Higgs particle (with vanishing Higgs condensate one-point function)
arising an inexorable companion of the interacting massive vectormeson by
consistency rather than by a free choice in model building. In fact we should
be able to see the true frontiers resulting from the physical principles
instead of those set by short distance problems and power counting of
Lagrangian quantization. The aforementioned idea of linking the crossing
property to the KMS property of an auxiliary on-shell QFT whose thermal
correlation functions describe the connected formfactors of a masterfield
should be seen as a first attempt at understanding the constructive side of
the uniqueness theorem\footnote{Any method by which one can compute
formfactors without going through correlation functions of pointlike fields in
intermediate steps is automatically ultraviolet-finite.}.

One reason (in addition to the strong return of QFT in the form of gauge
theories) why the old S-matrix approach was given up at the end of the 60s was
the lack of an operator formalism for the implementation of crossing; ideas
based on analyticity properties of formfactors using conjectured
non-perturbative properties (Mandelstam representations, and Regge poles) did
not lead to a framework in which one could systematically compute amplitudes,
just as the Kramers-Kronig dispersion theoretical framework of the 50s was
found unsuitable for systematic and trustworthy calculations (but at least
very well suited for experimental checks of consequences of causality).
Modular theory suggests that on-shell analyticity properties are deeply
related to domain properties of modular operators, and therefore it seems to
be more important for constructive attempts to unravel these operator
properties rather then using analyticity properties.

Physicists unlike mathematicians are not easily deterred by hard problems. If
they come across an important problem (as the Heisenberg-Chew S-matrix
attempts of a formulation of particle physics without ultraviolet
divergencies) they do not give up and leave it to the future because they are
unable to solve that problem by presently known methods; they rather tend to
invent another similar problem which is more susceptible to solution. In the
case at hand, Veneziano discovered that by using properties of $\Gamma
$-functions one can implement a property (called ``duality'') which formally
looks like the field theoretic crossing by replacing the scattering continuum
in the invariant energy of the elastic part of the S-matrix with a an infinite
discrete particle ``tower''. The reasons given at that time for tolerating
such an ad hoc looking structural violation of field-theoretic scattering
theory were purely phenomenological and are long forgotten in the modern
community of string theorist.

Veneziano's dual model construction based on properties of $\Gamma$-function
would probably also have been forgotten if it was not for two important
observations about two intriguing aspects. The first property was that the
model allows an alternative description in terms of an auxiliary d=1+1
conformal field theory which in some sense gave more respectability than
$\Gamma$-function engineering. The crossing property of the infinite particle
tower (which we will refer to as Veneziano ``duality'' in order not to confuse
it with the crossing property of a field theoretic S-matrix) was deferred to
the duality property in chiral conformal 4-point functions which is basically
the commutation structure (with braid group commutation relations in the
general case) which relates the various permuted positions of the conformal
operators\footnote{The tower structure arises from operator product expansions
between pairs of operators in the 4-point function.}. More important was the
later realization that models of this kind permit an interpretation in terms
of the quantization of a classical string described by a Nambu-Goto
Lagrangian. The required Poincar\'{e} invariance of the momentum space string
led to covariant realizations in 26, and in the presence of supersymmetry in
10 spacetime dimensions; what appeared initially as a vice was (in the string
theoretic credo) converted into a virtue by the invention of compactifying
spatial dimensions in the spirit of Kaluza and Klein. The Nambu-Goto string
quantization requires to picture the range of Veneziano's auxiliary chiral
fields as a target space of a first quantized map defined by a classical
reading of the chiral fields. This in turn introduces a rich differential
geometric structure which the auxiliary chiral reading of the old dual model
did not have. Therefore it is not surprising that the Green-Schwarz-Witten
formulation of string theory \cite{GSW} and its subsequent extensions has had
a significant impact on mathematics; however there are no indications from
experimental high energy particle physics that nature realizes such structures.

Since it is not the duty of a theoretician to worry about the status of the
experimental situation but rather to support or criticize ideas by conceptual
and mathematical analysis, the relevant question about this kind of string
theory in the present context is how it relates to contemporary QFT. This
boils down to the question whether the above duality property which replaced
the field theoretic crossing has destroyed causality and localizability and if
this turns out to be the case whether string theory offers a generalization of
these structures. Ever since the time of Jordan's quantization of wave fields
causality and localization structures are the most indispensable properties in
quantum physics\footnote{Recent attempts to formulate so-called
``Noncommutative Field Theories'' have shown that although it is
mathematically not difficult to abandon causality and localization, the
construction of a generalized substitute which still permits a physical
interpretation is an extremely difficult unsolved problem.} which are directly
linked to the physical interpretability of a theory. Without having a
localization structure it is not possible to interpret structural properties
in momentum space; particle momenta are primarily data which parametrize
relations between asymptotic spacetime events \cite{Haag}, covariance alone is
of no help. This is no problem for those natural quantum strings (e.g. Wigner
massless strings) in the setting of LQP. These strings exist because the
observable net in certain models allows ``charged'' representations and the
superselected charges have internal degrees of freedom which cannot be dumped
into pointlike field multiplets. Unlike Nambu-Goto strings these quantum
strings have an internal structure which makes them ``stringy'' in the
interpretable sense of semiinfinite string localization; the fact that they do
not permit a ``quantization reading'' makes them more noncommutative than
pointlike objects. There is no intrinsic physical meaning to quantum closed
strings; they would be the analogs of particle-antiparticles pair objects and
have the same charge as the vacuum. Since there is no Lagrangian formulation,
the implementation of interactions cannot follow the standard pattern of
causal perturbation theory (Wick-ordered couplings etc.).

In the standard setting, the quantum fields which commute for spacelike
distances with a given causal field (local equivalence class or ``Borchers
class'' of the given local field) follow a classical pattern, i.e. are local
composites (normal products which reduce to Wick products in case of free
fields). This relic of a classical picture is lost for those zero mass Wigner
objects which have an internal structure. In that case the class of relatively
local objects in the sense of spacelike distances is much larger than the
classically suggested local composites. The mathematically simplest
illustration of this new phenomenon is provided by generalized free fields
which have a continuous (not denumerable) class of relative local objects
\cite{Due-Reh}.

LQP has a very elegant way of dealing with this new phenomenon of extension
from the classically suggested picture of composites to that required by the
algebraic notion of relative spacelike commutativity. The algebraic extension
is called ``Haag dualization''. In the standard setting the operator algebras
which are localized in double cones generated by pointlike fields in the
vacuum representation are automatically ``Haag dual'' i.e. the commutant of
the spacelike disjoint algebra is exactly equal to the original operator
algebra. Haag dualization is a well-defined ``maximalization'' of local
algebras which at the beginning are not dual in this sense; the Hilbert space
remains unchanged in this operation. The Bisognano-Wichmann property implies
Haag duality for wedges, and the double cone algebras obtained from
intersecting Haag dual wedge algebras inherit this property. A new
perturbative formalism which replaces the causal perturbation theory of the
standard situation does not yet exist, but with the powerful guidance provided
by the algebraic setting it should be possible to construct one. Such a new
perturbative setting could be useful for implementing interactions for pure
quantum objects which do admit a (Lagrangian) quantization interpretation as
e.g. the aforementioned Wigner strings. Although they are still subject to
those causality and localization principles which Jordan discovered together
with his quantization of wave fields, they constitute generalized realizations
for which his 1929 Kharkov credo of ``abandonment of classical crutches'' is
much more than an esthetical improvement. For renormalizable interactions in
terms of pointlike fields the hypothetical on-shell procedure should reproduce
the renormalized mass-shell formfactors in a finite manner.

In the closed string theory based in the Nambu-Goto Lagrangian quantization
setting, the interactions are implemented by geometrical pictures of combining
and splitting tubes; however not only do these strings suffer from the severe
spacetime dimensional restrictions, but one has lost the relations with
causality and localizability, so that the physical interpretation becomes
obscure. In fact even if one imposes a multi-string tensor product structure
(``free string field theory''),\ the causality/localizabilty issue looks still
contradictory: in the d=24+1 lightfront setting the string becomes
``transparent'' \cite{Dimock1} (i.e. the algebraic commutation structure only
shows the center and not the string itself), whereas in the d=25+1 covariant
setting localization aspects seem to have been completely lost \cite{Dimock2}.
These calculations also reveal a fundamental difference in philosophy behind
LQP quantum strings and (Nambu-Goto) strings of string theory. Whereas LQP
views causality of observables and the resulting localization of
charge-carrying fields (point- or semiinfinite string-like) as the physically
indispensable properties\footnote{Momentum space properties acquire their
physical interpretation through localization of spacetime events; e.g.
particle momenta parametrize asymptotic localization events \cite{Haag}.} and
the covariance aspects as consequences (the connection being provided by
modular theory), string theorist place more emphasis on target space
covariance and seem to be less concerned about localization properties.

As far as the achievements of the quantized Nambu-Goto strings are concerned,
namely the incorporation of spin $s=2$ objects (``quantum gravity'') and the
ultraviolet finiteness, there is no good reason to think that pure quantum
strings (as the massless strings appearing in the Wigner zero mass
representation theory) are doing worse on these issues. In fact the idea that
the ultraviolet-finite formulation of the formfactor approach for factorizing
models may possess a general higher-dimensional counterpart was the present
Leitmotiv for what was considered as Jordan's legacy. Any approach which
either succeeds to calculate directly in terms of algebras without the use of
pointlike fields, or only uses only formfactors of fields and avoids
correlation functions will be ultraviolet-finite and the uniqueness theorem of
the inverse scattering problem \cite{S3} suggests that such a formulation exists.

The development of concepts starting from Jordan's quantized wave fields
passing through renormalization theory to the more intrinsic setting of LQP
has been a historic dialectic process in which new ideas were confronted with
established principles in an extremely profound and conscientious manner.
Since different from mathematics, theoretical physics at certain times has to
pass through very speculative phases, this dialectic confrontation (which
either eliminates the speculative idea as unfounded, or ends with a historical
synthesis which incorporates the old setting as a limiting case into a new
framework) is the most valuable theoretical tool which separates genuine
revolutionary enlargements of knowledge about nature from one way trips into
the ``blue yonder''. The list of abandoned proposals is much larger than that
of successful theories.

The present LQP setting, which is deeply rooted in the history of QFT by
elevating the Jordan-Pauli causality \cite{P-J}\cite{cau} combined with the
abandonment of ``classical crutches'', would view the string proposal based on
the quantization of the Nambu-Goto Lagrangian (which leads to high spacetime
dimensions and obscure localization aspects) as a step into the physical
``blue yonder''.

String theorists argue that their model of closed strings approaches quantum
field theory in the low energy limit \ But for making such arguments involving
scale-sliding on the formal level of actions more credible, one would need
some form of localization on the string side. A more convincing argument would
consist in showing that the dual S-matrix obtained from the string
prescription approaches a S-matrix with the field theoretic crossing property,
but the lack of conceptual understanding about the intrinsic meaning of the
string recipes make this a hopeless task. The replacement of time-dependent
LSZ scattering theory as a consequence of QFT locality by a string cooking
recipe without any hope of regaining the old conceptual elegance is another
worrisome aspect.

An even greater stumbling block for a conceptually conscientious quantum
physicist is the way string theorists implement interactions between their
first quantized strings by joining and splitting tubes. This picture does not
only return to pre LSZ times, but it even falls back behind Heisenberg's
dictum that the position of an electron is not an attribute of the electron
but of the event produced by the interaction with the measurement apparatus.
By not separating algebraic properties of observables from those which
originate from states, one is returning to a conceptual level of the
pre-Heisenberg Bohr-Sommerfeld quantum theory in which such a separation was
not possible.

String computations are expected to produce a dual S-matrix in the form of an
infinite series in ascending genii of Riemann surfaces. Instead of causality
and localization properties the construction of the dual string S-matrix uses
geometrical information about Riemann surfaces, Teichmueller spaces etc. But
there is not the slightest indication that these geometric properties are
intrinsic. Let us explain carefully what is meant by a property being
intrinsic. In the standard formulation of QFT one computes a system of
correlation functions of pointlike fields. A reconstruction theorem asserts
\cite{S-W} that all physical properties which went into the construction of
these correlations can also be re-extracted from these correlations; i.e. if I
pass this calculated system of correlations to my learned particle theory
colleague, he will be able to reconstruct a Hilbert space with a system of
causal and localizable operator algebras and verify all physical (local
gauge-invariant) properties without my telling him in what way I computed my
correlation functions; they are intrinsic properties of my calculated set of
correlation functions. Examples for non-intrinsic properties are those
properties which enter on the classical side if one started the computation
with ''classical crutches'' as Lagrangians, actions in functional integrals
etc.. Quantization is a form of art outside the range of a reconstruction theorem.

It is clear that reconstructability and intrinsicness are important conceptual
achievements of QFT and the uniqueness of the inverse scattering problem (in
case of asymptotically completeness) attributes an equally important role to a
field theoretic crossing symmetric S-matrix. In string theory no such
reconstruction theorem is known and none of those geometric properties which
form the basis of string theory have been shown to be intrinsic; in fact their
proximity to quantization properties casts serious doubts about their
intrinsic nature.

Whereas temporary shortcomings are sometimes unavoidable at the start of new
ideas, it is much harder to feel comfortable if after more than 20 years the
unclear situation prevails and the historical awareness gets lost. It is not
surprising but deeply lamentable, that the sociological dominance of string
theory threatens to destroy the conceptual fabric of one of the most
successful scientific endeavors which started with Jordan's quantization of
wave fields almost 80 years ago. It is particularly worrisome that profound
knowledge, which one expects to be important in the future development of
particle physics, gets lost at an alarming rate\footnote{Most string theorists
in the younger generation know what a Calabi-Yao manifold is, but only a few
have a rudimentary knowledge of LSZ scattering theory.}.

Within the historical context it would have been more natural to study first
those pure quantum strings (as e.g. the Wigner helicity tower representations)
and their interactions since they have deep roots in the causality and
localization principles. But generalizations along a given quantization
formalism are often much simpler than extensions which are subject to physical
principles for which a suitable formalism still needs to be developed. In the
apparent clash between the time-honored causality and localization ideas which
found their first expression in Jordan's work on quantization of wave fields
(and for which the same author was already looking for a more intrinsic
formulation free of classical ``crutches'') and the quantization-based
Nambu-Goto strings and their geometric generalizations, I have opted for the
path which saves the principles and sacrifices the quantization formalism.
Therefore it should not surprise that I view the encouragements of
experimentalizers by string theorists to detect the little curled up spatial
poltergeist dimensions which account for the difference between the numbers 10
and 4 of the supersymmetrized quantized Nambu-Goto string) with a certain
amount of suspicion and disbelief. Expensive experiments based on flimsy
theory are the most visible signs of crisis in particle physics. One can only
hope that this kind of experimental enterprises will not just end with a
negative result and without any other theoretically unexpected observation,
because this will damage more than the faith in the value of the
string-oriented post electro-weak particle physics.

If string theory would be a just a fashionable mathematically challenging area
which attracts young people and competes with other ideas, there would be no
reason to be worried. Even if one does not accept the arguments in favor of a
quantization interpretation\footnote{In order to talk about a geometric target
range, it would be necessary to be able to view the chiral conformal theory as
the result of quantization of a classical field structure.} of the range of
conformal fields as a string target space, the mathematical results of string
theorists on subtle problems in conformal field theories (e.g. the intricate
structure behind the Liouville field theory) are valuable extensions of
knowledge about nonperturbative QFT. The worrisome negative aspects are not
primarily coming from the scientific content, but rather from the sociological
climate and the hegemonial ideological stance of its meanwhile totally
globalized and omnipotent community. This is even visible in the terminology.
Whereas the aim of natural sciences for centuries was the de-mystification of
nature (essentially since the time of Lavoisier, when the phlogiston theory
was abandoned), string theorist do not miss any opportunity to emphasize that
their big Latin M could also be understood as ``Mystery'' and one does not get
the impression that they are interested in changing this situation. At the
time of the great quantum theoretical discoveries when the number of
physicists was much smaller than presently, there was no danger of the
scientific content being dominated by particular interests of particular
schools outside the quest for knowledge. The various physics institutes
existed in a multilateral equilibrium and a sociological amplification of an
idea leading to a monoculture of a theory without experimental support would
have been unimaginable ever since alchemy was abandoned. The large number of
physicists and the globalization through instant communication have made this
system very unstable. Whereas fashions in earlier times influenced preferences
in the selection of candidates mainly on a local level, the globalized trend
of the string ideology is placing a severe strain on the liberty of research.

We are accustomed to think of the exact sciences as an area which unfolds in a
more detached and objective way than other human endeavors. But there is no
rule which states that the exact sciences are immune against those human
shortcomings and defects which led to catastrophic political ideologies.
Physics was successfully defending its independence against infringements of
political ideologies; attempts in this direction ended as a lip service in
forewords of textbooks from the USSR, but were never able to influence the
scientific content. There is however no defense against ideological strands
which originates inside physics. Their power of control over minds and money
does not depend on whether they were able to contribute anything of enduring value.

The start of string theory dominance occurred in the 70s when the dual model
theory of strong interactions was converted to a theory of quantum gravity
(the Paris ``Bartholomew night-like massacre of the old string theory''). One
would have expected that a theory after having had some success at laboratory
energies in strong interactions would loose its physical credibility if one is
asked to abandon this application and instead use that theory as a theory of
gravity at energies corresponding to the Planck length which are 15 orders of
magnitudes higher. But it is also clear that in case one succeeds to convince
a majority that this is a reasonable step, one has created a new area of
research which different from the original strong interaction theory is only
controlled by mathematics and outside obligations of experimental
verifications. This new situation has led to the present state in which
particle physicists solving string theoretic problems seem to be doing quite
well, but particle physics is in a deep crisis (the particle physics analog of
the economic Enron-Worldcom effect). Physics and what physicists are doing is
presently not necessarily identical.

It may be seen as a historical irony that Germany as the country where most of
it began almost 80 years ago, is also the place where this trend of a string
oriented particle physics monoculture has most progressed. Research proposals
in particle theory which does not relate in some way with strings have no
chance of being approved, even if their scientific content is of the highest
quality. The necessity of getting the string imprimatur is an insult to free
research. The presence of physicists in leading positions of influence who
have since their graduate students times not done anything else but string
theory and who built their career on their social success in that area (there
was yet no scientific success), is a heavy liability for the future of
particle physics. Social success tends to destroy the intellectual modesty
which one needs to start something new and original. It also diminishes the
intellectual curiosity and ability to appreciate different ideas in particle
physics and causes an amnesia regarding historical roots and continuity. Since
there is not a single pure research institution dedicated to particle physics
in Germany whose future is not compromised in this direction, this may very
well be the end of the project which I have described in this essay as
Jordan's legacy; this is the reason why in the present context one cannot be
silent about this very serious problem.

There have been attempts by philosophers and historians to understand
important discoveries in the context of their sociological settings. The 1925
discovery of quantum mechanics with its rejection of the causality in the
sense of classical determinism has been related to the post World War I gloom
and doom of a nation which only a short time before was thinking of itself as
one of the intellectual centers of the world; i.e. playing with acausality as
a way to overcome the gloom and stay in the intellectual avant-garde through a
revolutionary act \cite{Form}\cite{Schweber}. People who subscribe to such
ideas may perhaps ask the question whether string theory and in particular
those more recent proposals which come with big Latin letters as M-theory
should be considered as the sociological manifestation of the Zeitgeist of the
Hegemon in particle physics.

We have used some writings of Pascual Jordan as one of the protagonists of
quantum field theory as a historical support of a new program which
de-emphasizes pointlike fields in order to obtain an intrinsic
singular-coordinate free formulation which bypasses the ultraviolet aspect of
the standard formulation. One may receive additional support in this endeavor
from his glorious friendly opponent Paul Dirac. In most of recent centennial
articles his strong believe in mathematical beauty and conceptual harmony is
emphasized and as an illustration his beautiful geometric presentation of the
Dirac equation and construction of magnetic monopoles are cited. But the same
Dirac also rejected renormalization theory, a fact which is mostly ignored as
an unpleasant shortcoming of an otherwise brilliant theoretical physicist. It
is my firm conviction that a man who contributed so much to the beginning of
renormalization theory certainly did not reject its impressive results out of
a lack of understanding, but rather because he considered it as conceptually
unfinished and mathematically grossly imperfect. Indeed if one looks at
standard approaches as canonical quantization or quantization via functional
integrals, one realizes that this is what deserves to be called ``as if''
physics. By this I mean that the physical correlation functions (i.e. the
results after renormalization) are Einstein-causal but definitely not
canonical nor representable in terms of a functional integral. The ``as if
(everything would be consistent)'' attitude (i.e. the ignoring of the mismatch
between the original assumptions and the final mathematical structure /which
one arrives at after mathematical doubtful intermediate steps) generated a
false impression of a universality of the functional framework. The majority
of particle physicists ignore this imbalance and some even may react irritated
against anybody who points his finger at this serious imperfection. The ``as
if'' aspect is often unavoidable with new discoveries; it is important to
first secure the results and later worry about the mathematical
legitimization. There is however nothing more contraproductive in theoretical
physics than a solidified ``as if scheme'' which has been used for decades by
generations of physicists since the interest to discover the mathematical
correct way (which may require new conceptual investments far beyond the ``as
if scheme'') decreases with time.

My explanation for Dirac's rejection of renormalization is that although he
did not doubt the validity perturbative results, the ``as if'' attitude in
their derivation and the mathematically delicate setting in terms of singular
field coordinatizations was not in line with his sense of beauty. In fact he
may even have rejected the mathematically improved setting of causal
perturbation theory\footnote{The causal perturbation formulation is also more
general than the Euclidean action approach in that no free Lagrangian
description (Euler-Lagrange equation of motion) is required. Most
covariantizations of Wigner's representation theory \cite{Wein} lead to free
field coordinatizations which do not permit a Lagrangian description.} because
that more rigorous approach draws a technical frontier between
renormalizable/nonrenormalizable without giving a hint about its relation
relative to the underlying physical principles (there exists no dictum that
one must implement interactions by coupling free fields, although it is
difficult to think of anything else if one starts with Jordan's quantized wave
fields). Unfortunately the ``as if'' attitude in physics which was totally
absent in Jordan's times (and is still absent in LQP) pervades most formalisms
used in particle theory, it is in particular characteristic of string theory
whose intrinsic physical content is unknown.

The LQP setting enriched with modular theory, advocated here for constructive
purposes, may be revolutionary enough to have satisfied Jordan's demands, and
its conceptually concise setting may even have lived up to Dirac's
expectations concerning beauty and rigor, but the main question: \textit{is
the success of this method in the case of d=1+1 factorizing model a strange
accident or is there a general message behind, }despite many encouraging
arguments, has to be left open. However the many new profound questions which
arise in the present context show that despite its more than 75 years of
existence after its discovery in 1927, QFT is very much alive and its deeper
conceptual and mathematical layers still need to be understood in agreement
with Jordan's expectations at the 1929 Kharkov conference. When string
theorist these days point at old (and in their view understood) QFT as opposed
to their new string theory, they tacitly identify QFT with its very limited
Lagrangian implementation.

Acknowledgments: I owe thanks to Engelbert Sch\"{u}cking for informations
concerning previous article and books related to Jordan. I am particularly
indebted to Anita Ehlers for making her unpublished biographical notes
available to me. Finally I thank J\"{u}rgen Ehlers for several interesting
discussions and help on some problems of interpretations in Jordan's papers
which required an understanding within the Zeitgeist of that epoch. I am also
indebted to Stanley Deser and William Brewer for valuable suggestions of improvements.

\bigskip\ 
\end{document}